\documentclass{aastex}

\usepackage{color} 

\usepackage{graphicx}
\graphicspath{ {pdf_figs/} }
\usepackage{amssymb}
\usepackage{amsmath}
\usepackage{epstopdf}

\shorttitle{Photoevaporated Disks with Non-Uniform Alpha} 
\shortauthors{Kalyaan et al.}

\begin{document}

\title{External Photoevaporation of the Solar Nebula II: Effects on Disk Structure and Evolution with Non-Uniform Turbulent Viscosity due to the Magnetorotational Instability}

\author{A.~Kalyaan}
\affil{School of Earth and Space Exploration, Arizona State University, PO Box 871404, Tempe, AZ 85287-1404}
\email{anusha.kalyaan@asu.edu} 

\and 

\author{S.~J.~Desch}
\affil{School of Earth and Space Exploration, Arizona State University, PO Box 871404, Tempe, AZ 85287-1404}

\and 

\author{N.~Monga}
\affil{Department of Physics, Arizona State University, PO Box 871504, Tempe, AZ 85287-1404}



\begin{abstract}
The structure and evolution of protoplanetary disks, especially the radial flows of gas through them, are sensitive to a number of factors. One that has been considered only occasionally in the literature is external photoevaporation by far-ultraviolet (FUV) radiation from nearby, massive stars, despite the fact that nearly half of disks will experience photoevaporation. Another effect apparently not considered in the literature is a spatially and temporally varying value of $\alpha$ in the disk [where the turbulent viscosity $\nu$ is $\alpha$ times the sound speed $C$ times the disk scale height $H$].  Here we use the formulation of Bai \& Stone (2011) to relate $\alpha$ to the ionization fraction in the disk, assuming turbulent transport of angular momentum is due to the magnetorotational instability. We calculate the ionization fraction of the disk gas under various assumptions about ionization sources and dust grain properties.  Disk evolution is most sensitive to the surface area of dust.  We find that typically $\alpha \lesssim 10^{-5}$ in the inner disk ($< 2$ AU), rising to $\sim 10^{-1}$ beyond 20 AU.  This drastically alters the structure of the disk and the flow of mass through it: while the outer disk rapidly viscously spreads, the inner disk hardly evolves; this leads to a steep surface density profile ($\Sigma \propto r^{-\langle p\rangle}$ with $\langle p\rangle \approx 2 - 5$ in the 5-30 AU region) that is made steeper by external photoevaporation. We also find that the combination of variable $\alpha$ and external photoevaporation eventually causes gas as close as 3 AU, previously accreting inward, to be drawn outward to the photoevaporated outer edge of the disk. These effects have drastic consequences for planet formation and volatile transport in protoplanetary disks.
\end{abstract}


\section{Introduction}

Protoplanetary disks form around low mass stars as a consequence of stellar formation when collapse of a slowly spinning molecular cloud core transforms it into a rapidly rotating star-disk system. Once formed, the disk undergoes viscous evolution via shearing stresses that are set up through differential rotation of its gas and dust constituents. Much of the disk mass flows inward and accretes onto the star, while simultaneously a portion of its mass is transported far outward (to conserve angular momentum), causing the disk to continuously spread outwards throughout its $\sim$ 10 Myr evolution (Pringle 1981). Planetesimals and planetary cores form via coagulation and accretion of the remaining dust and gas on timescales of a few Myr or less, before the disk is dissipated. All of these processes, and the mass available for planet formation, rely on the details of how matter is moved through protoplanetary disks. The surface density profile $\Sigma (r)$  - or mass per area of disk as a function of distance from the star $r$  - determines how much of mass might have been available in the feeding regions of the planets; its evolution over time determines how mass moves in the disk and also how the planetary masses grow. 
Theoretical models of disk evolution are based on the canonical equations by Lynden-Bell \& Pringle (1974; hereafter LBP). These models assume that the disk evolves via shearing stresses mediated by a turbulent viscosity  $\nu$ that varies as $\nu$ $\propto$ $r^{\gamma}$ where $\gamma$ $\sim$ 1. They predict that $\Sigma (r,t)$ should approximate a power law $\Sigma(r) = \Sigma_0(r/r_0)^{-p}$ across much of the disk, with slope $p$ $\sim$ 1. Model predictions of disk integrated properties appear consistent with observations of disks in low mass star forming regions such as Taurus (Hartmann et al.\ 1998). Observations of resolved disks, on the other hand, can provide direct estimates of $\Sigma(r)$. Recent work with millimeter wavelength surveys of disks in another low-mass star forming region, Ophiuchus (Andrews et al.\ 2009; Andrews et al.\ 2010), fit $\Sigma(r)$ to a profile $\Sigma$  $\propto (R / R_c)^{-\gamma} \, {\rm exp}[-(R/R_c)^{2-\gamma}]$ ($R_c$ is the characteristic radius where the shape of the $\Sigma$ profile changes from the power-law to the exponential taper) which is similar to the LBP similarity solutions, and find $\gamma$ to be within the range 0.4 - 1.1, with a median value of about 0.9. This seems to be consistent with theoretical predictions of $\Sigma(r)$ of viscously evolving disks.
Despite this tentative match, one must use caution when inferring the distribution of mass from such observations. It is not even certain that all the mass is being observed, as several factors may lead to the disk mass in any annulus being underestimated.  The millimeter opacity of solids is sensitive to changes in grain size and composition (Beckwith et al.\ 1990, Beckwith \& Sargent 1991).  Converting a solids mass to a mass of gas requires knowledge of the uncertain dust-to-gas mass ratio. Also, some massive disks may still be optically thick in the sub-mm regime, hence shadowing some of the disk mass (Andrews \& Williams 2005). Sub-mm observations are also not sensitive to $>$mm-sized dust grains, and hence may not account for mass locked up in larger grains, or even planets that have already formed, as they will remain undetectable for several Myr. Thus it is difficult to definitively derive $\Sigma(r)$ from astronomical observations.
 
Although the Sun's protoplanetary disk has long ago dissipated, an estimate of $\Sigma(r)$ (in a snapshot or time-averaged sense) can be obtained from the known masses and compositions of the planets. Weidenschilling (1977b) and later, Hayashi (1981) developed the so-called Minimum Mass Solar Nebula (MMSN) model in which an estimate of the surface density profile of the solar nebula is found by augmenting the known mass of each planet (located in its present day orbit) with H$_2$/He gas to bring it to solar composition, and then dividing this augmented mass by the area of the annulus in which it orbits. An estimate of $\Sigma(r)$ is found at each planet's radial location $r$, and a power law can be fit to these points. A widely used equation for the MMSN model put forward by Hayashi (1981) is :
\begin{equation}
\Sigma(r) =1700 \left( \frac{r}{1\, {\rm AU}} \right)^{-3/2}  \,{\rm g} \,{\rm cm}^{-2}  .
\end{equation}
Later, this model was extended to extrasolar planetary systems as the Minimum Mass Extrasolar Nebula (MMEN) model (Kuchner 2004; Chiang \& Laughlin 2013; and Raymond \& Cossou 2014), using some of the $>$ 470 known multiple planet systems. For such close-in planetary systems, $\Sigma(r)$ can only be inferred up to a few AU. Chiang \& Laughlin (2013) find the slope of $\Sigma (r)$ to be $p$ $\sim$ 1.6 - 1.8, while Raymond \& Cossou (2014) argue that $\Sigma(r)$ varies wildly amongst planetary systems and $\sim$ 1.6 is only a median value. With future data from more widely separated planetary systems, any universal or median MMEN will provide for much better comparison with the MMSN model than current data. The slopes for $\Sigma(r)$ inferred from the MMSN and MMEN models, $p$ $\approx$ 1.5-1.6, are steeper than the slope of the profile measured from observations or predicted by theory. However, the MMSN and MMEN models suffer from many shortcomings. For the MMEN model, there are large uncertainties or lack of data on mass, radii and the composition of planets, and drawing out a surface density profile mandates assuming a uniform (usually chondritic) composition, and usually a mass from a known radius assuming a mass-radius relation (Chiang \& Laughlin 2013). It also inherently assumes that the planets were formed where they are now observed. As for the MMSN, while it offers a direct reference measurement of $\Sigma(r)$ from our own solar system, it only accounts for the minimum amount of mass in the solar nebula that was sequestered into the final planets. It assumes that no solids were lost from the nebula throughout its evolution and also only samples the disk at one given point in nebular history, i.e., after the outer planets assumed their final positions in their current orbits. Both the MMSN and the MMEN models do not account for the migration of planets in the disk, when numerous observations of close-in hot massive planets in exoplanetary systems suggest significant planetary migration driven by exchange of angular momentum with the disk gas as well as planetesimals (Armitage 2007; Crida 2007; Walsh et al.\ 2011; Kley \& Nelson 2012). Planetary migration, if present, can later dramatically modify the initial surface density profile that was available for planet formation.

It was in this context that Desch (2007) argued that the dynamical constraints from the Nice Model (Tsiganis et al.\ 2005; Gomes et al.\ 2005; Morbidelli et al.\ 2005) at $\sim$ 880 Myr after the formation of the disk provide for an MMSN model better suited for studying the structure of the early disk. The Nice Model argues that the giant planets are likely to have formed from a more compact configuration, at 5.45, 8.18, 11.5 and 14.2 AU for Jupiter, Saturn, Neptune and Uranus respectively (in which the two ice giants likely swapped places). Substantial migration of outer planets eventually led to their final positions today, spread across 5-30 AU. With this configuration, the Nice Model successfully explains many dynamical constraints of the solar system including the observed orbital parameters of the giant planets, as well as the halt of Neptune's migration, dynamical classes of the Kuiper Belt, origin of the Jovian Trojan asteroids and the Late Heavy Bombardment. Desch (2007) used these updated positions of the giant planets in the MMSN model to find a $\Sigma(r)$ profile that was steeper than the MMSN model with a slope of  $p \sim 2.2$:
\begin{equation}
\Sigma(r) =343 \left (\frac{f_p}{0.5} \right)^{-1} \left (\frac{r}{10\, {\rm AU}} \right)^{-2.168} \,{\rm g} \,{\rm cm}^{-2}  ,
\end{equation}
where $f_p$ is a factor describing the fraction of solids in dust at the end of planet formation. This compact architecture results in a higher $\Sigma(r)$ throughout the disk and also in a steeper $\Sigma$ profile. Desch (2007) found that such a steep profile matches very well with the solution of a steady-state \textit{decretion disk} (Lee et al.\ 1991) i.e., a disk which is losing mass radially \textit{outward}. He argued that this disk mass loss process can be explained very well by photoevaporation due to intense far ultraviolet (FUV) radiation from a nearby massive star.

Photoevaporation is an efficient mechanism for disk dispersal, in which the disk is impinged by external extreme-ultraviolet (EUV) or FUV radiation that causes the gas in the upper atmosphere of the disk to heat to $\sim$ $10^{2}\,{\rm K}$ and escape the gravitational potential of the star. It was directly observed in the Trapezium cluster of the Orion Nebula where disks close to $\theta^1$ Ori C (the O star in the Orion Nebula) were not only found to be truncated (McCaughrean \& O$'$Dell 1996) but also were observed to be losing mass steadily with mass-loss rates of up to  $\sim 10^{-7}$ $\rm{M_{\odot}\,yr^{-1}}$ (Henney \& O$'$Dell 1999). Recent Atacama Large Millimeter Array (ALMA) observations by Mann et al.\ (2014) rule out any observational bias and confirm a distinct lack of massive disks close to the O star ($\sim$ 0.03 pc). Disks born in low-mass star forming regions (like Taurus or Ophiuchus) however, viscously spread to large radii $\sim$ 300 AU (Hartmann et al.\ 1998; Andrews \& Williams 2007; Andrews et al.\ 2009, 2010), in contrast to truncated photoevaporated disks.

It is very likely that the Sun's protoplanetary disk experienced photoevaporation. From observations of nearby clusters, it is expected that roughly 50\% of all disks are likely to be present in intensely irradiated birth environments with sufficient FUV flux to cause significant mass loss via external photoevaporation (Lada \& Lada 2003).  The abundances of short-lived radionuclides that are more likely to have been created in nearby supernovae (Wadhwa et al.\ 2007) and then injected into the solar nebula (Ouellette et al.\ 2010; Pan et al.\ 2012) very likely betray the presence of nearby massive stars. The orbit of Sedna also suggests that it is likely to have been perturbed inward into the solar system due to a nearby passing star (Kenyon \& Bromley 2004). The edge of the Kuiper Belt at $\sim$47 AU is also consistent with a disk that is seemingly truncated in a clustered environment, which could be attributed to either cluster dynamics where a passing star strips material off of the disk (Clarke \& Pringle 1993; Kobayashi \& Ida 2001; Adams 2010) or truncation due to photoevaporation (Trujillo \& Brown 2001; Hollenbach \& Adams 2004). Finally, the oxygen isotope anomalies found in Ca-Al inclusions in meteorites is likely resolved by an isotopically selective fractionation caused by the self-shielding of CO against photodissociation by external FUV radiation (Lyons et al. 2009). It has also been argued that external FUV radiation is also likely to create enormous quantities of amorphous ice in the cold outer disk (Ciesla 2014; Monga \& Desch 2015) that will be able to trap noble gases, which upon radially migrating inward lead to the noble gas abundances measured by the Galileo mission in Jupiter's atmosphere (Monga \& Desch 2015). External photoevaporation therefore very likely affected the structure and dynamics within our protoplanetary disk.

If external photoevaporation affected the surface density profile of our protoplanetary disk in the manner predicted by Desch (2007), the outer parts of the solar nebula would be described as a steady-state decretion disk. Mass would flow from a reservoir in the inner disk, outward with a constant mass decretion rate $\dot{M}$ through the outer disk, to an outer edge where it is lost by photoevaporation. In the 5-30 AU region of the disk, a slope $p$ $\approx$ 2.2 is predicted. More recently, Mitchell \& Stewart (2010) performed numerical simulations of disks subjected to external photoevaporation to test whether the steady-state decretion disk solution of Desch (2007) applied. From their simulations, they report quasi-steady state disks with less steep profiles having slopes $p$ $\sim$ 1.6 - 1.8. This is not as steep as the profile inferred by Desch (2007), but the discrepancy may have to do with the way viscosity is handled.
The viscosity of Mitchell \& Stewart (2010) was parameterized in the usual way, with the fiducial $\alpha$ scaling relation from Shakura \& Sunyaev (1973) in which $\nu$ = $\alpha\,c_s H$, where $c_s$ is the speed of sound, which denotes the maximum velocity scale of turbulence; $H=\,c_s/\Omega$ is the scale height of the disk, which denotes the maximum size scale of turbulence assumed; and $\alpha$ is the dimensionless scaling factor for turbulent viscosity that represents the efficiency for angular momentum transport.  It is important to note that Mitchell \& Stewart (2010) assumed a uniform value for $\alpha$ throughout the radial extent of the disk. We assert that this is an ad hoc assumption unless an actual mechanism for angular momentum transport is identified. Likely mechanisms do not predict uniform $\alpha$. For example, one mechanism that has often been proposed for angular momentum transport  - at least early in the evolution of the disk, while it is still massive - is the gravitational instability (GI) in which $\alpha$ depends on the Toomre parameter $Q_T = c_s\,\Omega/\pi G\,\Sigma $ (here $\Omega$ is the orbital frequency) as given by the following prescription from Lin \& Pringle (1990) :
\begin{equation}
\alpha = 0.01 \left( \left[\frac{Q_{\rm crit}}{Q_{T}}\right]^2 - 1 \right) , 
\end{equation}
where $Q_{\rm crit}$ is the minimum value of $Q_T$ at which the disk becomes gravitationally unstable. Since $Q_T$ is clearly dependent on $r$, $\alpha$ can be expected to be variable through the radius of the disk, if disk viscosity originated due to GI. 
 
The most widely accepted mechanism attributed to the transport of mass and angular momentum across the disk is the magnetorotational instability, or MRI (Balbus \& Hawley 1998; Gammie 1996), whose operation is dominant in regions of the disk where gas with a sufficiently high ionization fraction is coupled to the magnetic field. A disk with a varying density profile across radius and height, ionized mainly by stellar X rays and cosmic radiation, would have an ionization fraction varying by several orders of magnitude. Such variation in ionization levels is also apparently observed in the TW Hya disk by Cleeves et al.\ (2015), which is possibly due to the spatially varying ability of stellar wind to repel cosmic rays (Cleeves et al.\ 2014). A varying ion fraction would also result in a variable $\alpha$ across the disk.  Dense gas and dust-rich protoplanetary disks are only partially ionized systems, and hence it is important to consider the effects of non-ideal magnetohydrodynamics (MHD) in the operation of the MRI. Ambipolar diffusion takes particular importance as it operates in the highly ionized and low density regime which primarily constitutes the disk atmosphere and large portions of the outer disk (Bai \& Stone 2011). We begin our simulations with disk mass $M_d$ = 0.1 $\rm{M_{\odot}}$ - the expected upper threshold for a gravitationally stable disk, and incorporate the formulation of Bai \& Stone (2011) for deriving MRI-viscosity from ionization state by including the non-ideal MHD effects of ambipolar diffusion, to estimate the value of $\alpha$ across the radial and vertical extent of the disk. A similar effort of including MRI derived viscosity with non-ideal MHD effects in PPD simulations was also undertaken by Landry et al.\ (2013). They perform simulations where they include ambipolar diffusion using the prescription of Bai \& Stone (2011) as well as ohmic resistivity, but do not discuss photoevaporation that can also significantly affect disk behavior. In this work, we argue and show that considering a variable value for $\alpha$ in disk models can significantly affect the steepness of the disk profile. We also simultaneously incorporate external photoevaporation (important for the Sun's disk) in our models and show how disk structure and evolution are dramatically altered by considering both non-uniform $\alpha$ and external photoevaporation.

Finally, gas and dust grain chemistry play a vital role in determining the ionization fraction in each region of the disk. Previous works have employed: i) simple dust models that include only a single ion-based chemistry (molecular ion- or metal ion-based) adapting work from Oppenheimer \& Dalgarno (1974) (e.g. Fromang et al.\ 2002); ii) more complex chemical networks that account for multiple interacting species (e.g. Sano et al.\ 2000; Ilgner \& Nelson 2006); and iii) reduced chemical networks that attempt to simplify the complex reaction networks into simpler networks for easier computation (e.g. Semenov et al.\ 2004). Ilgner \& Nelson (2006) present a comprehensive comparison between different models that are commonly used for dust chemistry in disks. We use a simple dust model based on the first approach: a single ion-based approach. We also vary different sources of ionization. We test different models that focus on chemistry of either metal atoms or molecular ions, and show how they each affect the structure of the disk. 

The following sections are organized as follows. Section 2 will describe the details of our numerical models, how we include the effects of non-uniform $\alpha$ and photoevaporation, as well as our dust chemistry model. Section 3 will discuss the main results of the time evolution of $\Sigma(r,t)$ from our simulations with reference to a canonical simulation with typical values of each variable parameter in our models. We will also describe the effect of variation of each parameter. Finally, in Section 4, we discuss in detail what implications our results have towards planet formation.

\section{Methods}

In this section, we will discuss the numerical model for disk evolution in detail. We will first describe the underlying viscous disk evolution code, then our implementation of ionization equilibrium with dust chemistry throughout the radial and vertical extent of the disk in order to estimate $\alpha$ from MRI-viscosity, and finally the treatment of external photoevaporation due to FUV radiation from a nearby massive star. To understand how the Sun's nebula might have probably evolved, it is important to include the effects of external photoevaporation into a non-uniform $\alpha$ viscosity disk evolution model.

\subsection{Viscous Disk Evolution}

Our `1.5-D' disk evolution code employs the fiducial equations of viscous disk evolution from LBP where the rate of change of surface density $\Sigma(r)$ is related to $\dot{M}(r)$, the rate of inward mass flow through an annulus of the disk:
\begin{equation}
 \frac{\partial \Sigma} { \partial t} = \frac{1}{2 \pi r} \frac{\partial  \dot{M} }{\partial r} \,,
\end{equation}
where
\begin{equation}
 \dot{M}=6 \pi r^{1/2} \frac{\partial}{\partial r} \left(r^{1/2}\Sigma \nu \right) .
\end{equation}
Here $\dot{M} > 0$ refers to an inward mass flow toward the star, while $\dot{M} < 0$ refers to an outward flow towards the disk edge. Equation 5 can also be written as :
\begin{equation}
\dot{M} = 3 \pi \Sigma \nu \left[ 1 + 2Q \right] ,
\end{equation}
where $Q = \partial\,\rm{ln} (\Sigma\nu)/\,\partial\, \rm{ln}\, \textit{r}$.
The above equations are discretized into a logarithmic grid of 60 radial zones split across 0.1 AU to 100 AU for all of our runs (excepting the uniform $\alpha$ runs where we use a 100 radial zones, instead of 60). These are explicitly integrated in time. Mass fluxes are considered at the boundaries of each annulus, while viscosity $\nu(r)$, surface density $\Sigma(r)$, density $\rho$, ion abundance $N_{i}$ and electron abundance $N_e$ are all considered at the midpoint of each annulus.

We implement an initial surface density profile at time $t=0$ from the LBP self-similar solutions (Hartmann 1998):
\begin{equation}
 \Sigma(r)= \frac{M_0}{2\pi R_0\,r} \exp \left( -\frac{r}{R_0}\right), 
 \end{equation}
where initial disk mass $M_0$ is assumed to be 0.1 $\rm{M_{\odot}}$, and $R_0$ denotes initial disk radius, assumed to be 100 AU. We assume the mass of the host star is 1 $\rm{M_{\odot}}$. The ratio of stellar mass to disk mass of  0.1 represents the typical value of a most massive disk that is likely gravitationally stable. We incorporate the temperature profile for a disk undergoing layered accretion from Lesniak \& Desch (2011), which is suitable for a passively heated disk or a disk heated by MRI-driven accretion with $\dot{M}\,<\,10^{-7}\,{\rm M_{\odot}\, yr^{-1}}$:
\begin{equation}
 T(r) = 100 \left(\frac{r}{{\rm 1\,AU}}\right)^{-0.5}  {\rm K}
\end{equation}
We allow the disk to extend freely out to an outer computational boundary $r_{\rm out}$ by assuming that at $r_{\rm out} \gg r_{\rm disk}$, $\dot{M}=0$. For the inner boundary, we assume the zero-torque boundary condition, assuming that gas at some point becomes coupled to the slowly rotating star and must orbit at less than Keplerian velocity. This forces  $\partial\,\Omega/\partial \,r\, > 0$ close to the star,  $\partial\,\Omega/\partial \,r\, = 0$ at some boundary, merging with Keplerian rotation with  $\partial\,\Omega/\partial \,r\, < 0$ beyond that boundary. The boundary is fixed to be close to the stellar radius, although magnetospheric truncation of the disk (Bouvier et al. 2007) could move the boundary outward. This may slightly alter the structure of the disk in the innermost few tenths of an AU but will not affect its evolution in the outer disk that is the focus of the present work. Our boundary criterion is derived from the following analytical solution of Equations 4 and 5:
\begin{equation}
 \Sigma(r)= \frac{\dot{M}}{3\pi \nu}\left[1-\left(\frac{r_0}{r}\right)^{1/2}\right],
\end{equation}
where a uniform $\dot{M}$ and a narrow first zone are assumed. We then solve for $Q$ (from Equation 6) for the first zone by integrating the analytical solution (Equation 9) with $r$ to obtain the total mass of the first zone. Dividing by the surface area of the first annulus, this $\Sigma_1$ is equated to the analytical solution to solve for Q, and thereafter $\dot{M_1}$. 
We have evolved all simulations for 10 Myr, except in the cases where photoevaporation dissipates most of the disk such that the radius of the disk is truncated to 5 AU or less within the simulation timescales. In these cases, the simulations are terminated when the size of the disk shrinks to $\leq$ 5 AU in radius. 

We incorporate the standard $\alpha$ parameterization for turbulent viscosity by Shakura \& Sunyaev (1973), where $\nu$ = $\alpha\,c_s H $ and $\alpha$ is the turbulent viscosity coefficient, $c_s$ is the sound speed and $H$ is the disk scale height. From measurements of disk masses and accretion rates in disks in Taurus and Chameleon I star forming regions ($\sim$ 1 Myr old), Hartmann et al.\ (1998) inferred a globally averaged $\alpha$ $\sim$ 0.01, the value most disk models use. More recent studies (Andrews et al.\ 2009, 2010) find a range of $\alpha$ $\approx$ 0.0005 - 0.08 in resolved disks in the $\sim$ 1 Myr old Ophiuchus star-forming regions. We emphasize that any realistic physical mechanism of angular momentum transport is not likely to yield a constant value of $\alpha$ throughout the radius of the disk and its lifetime. In the following subsection, we will describe how we incorporate a non-uniform value of $\alpha$ derived from MRI viscosity.

\subsection{Viscosity from Magnetorotational Instability}

\subsubsection {Ionization Equilibrium with Dust}

We divide each radial zone further into 25 vertical zones across the thickness of the disk, from the midplane to its surface, to estimate the ionization fraction across radius $r$ and height $z$ of the disk. Vertical zones are chosen with the help of weights and abscissa of the Gaussian-Legendre quadrature. We assume that the disk is isothermal with height, and hence incorporate a simple gaussian profile for density $\rho$ across the height of the disk centered on the mid-plane:
\begin{equation}
\rho(r,z)= \frac{1}{2} \,\rho_o(r) \exp\left( -\frac{z^2}{2H^2}\right) .
\end{equation}
We thereafter assume that the disk is ionized by two sources of non-thermal radiation: X rays from the central host star itself, and cosmic radiation. Cosmic radiation, less intense than X-ray radiation, impinges the disk equally throughout $r$. As a consequence it affects a large fraction of the optically thin outer disk as well as the disk surface layers closer to the star. For the cosmic ray ionization rate across $r$ and $z$, we incorporate the widely used expression for Galactic cosmic rays given by Umebayashi \& Nakano (1981):
\begin{equation}
\zeta_{\rm cr}(z)= 1 \times 10^{-17} \exp\left(\frac{-\sigma_{\perp}}{100 {\rm \,g \,cm^{-2}}} \right) \,\,\, {\rm s}^{-1}
\end{equation} 
Stellar X rays, on the other hand, will strongly illuminate and penetrate the innermost disk and the optically thin surface layers. The disk midplane regions are likely to be shadowed by the dense inner regions, but the outer flared disk will be illuminated by central star's X rays, although not as intensely as in the inner disk; the X-ray ionization rate reduces with $r$. We use the X-ray ionization rates of Glassgold et al.\ (1997), who consider an X-ray emitting region of size $\sim$ 10$\,R_{\odot}$ centered on the star :
\begin{equation}
 \zeta_{\rm xr}(z) = \zeta_0\, Z_0   \;\,\,\,\; {\rm s}^{-1} \,\, , 
\end{equation}
where $\zeta_0$ is given by
\begin{equation}
\zeta_0 (z)= 6.45\times 10^{10} \, \tilde{\sigma}\, \left(\frac{kT_{\rm xr}}{1\, \rm keV}\right)^{-n} \left(\frac{L_{\rm xr}}{1\times10^{29} {\rm \,ergs\,s^{-1}}}\right) \left( \frac{r_{\rm mid}}{1\, \rm AU}\right)^{-1} \,\cos \theta \;\;\,\,  {\rm s}^{-1}, 
\end{equation}
$\tilde{\sigma}=2.27 \times 10^{-22}\, {\rm cm}^2$ is the photoionization cross section at 1 keV, $T_{\rm xr}$ and $L_{\rm xr}$ are the X-ray temperature and X-ray luminosity, respectively. Here, $kT_{\rm xr}$ is assumed to be 5 keV and $n=2.485$. 
$Z_0(z)$ is given by
\begin{equation}
Z_0(z) = A \left[ \tau(z)\right]^{-a} \exp\left[-B \tau(z)^b\right],
\end{equation}
where the respective constants are $A = 0.800$, $B = 1.821$, $a = 0.57$ and $b = 0.287$ (Glassgold et al.\ 1997). For the above equations, as shown in Fig.\ 1 (in Glassgold et al. 1997) we define optical depth $\tau(z)=1$ where $\tau(z)<1$. For higher optical depths, we assume the following expression from Glassgold et al.\ (1997):
\begin{equation}
\tau(z) = 42.76\,  \sigma_{\perp}(z) \,\tilde{\sigma}\, \left(\frac{kT_{\rm xr}}{1\,\rm keV}\right)^{-n} \frac{1}{\cos \theta} ,
\end{equation}
where, $\sigma_{\perp}(z)$ is the surface density normal to the disk mid-plane at about a height $z$ ($\Sigma=2\,\sigma_{\perp}(z)/\sigma_{\rm \,total}$ ; similar to column mass density as a function of $z$). In the above equations, we have included a factor $\cos \theta$ to account for the disk flaring angle at each $r$, as the fraction of the disk that would actually intercept the stellar X rays at each $r$ is dependent on the flare angle, and is implemented here as follows.
The disk is divided into two regions: the first is an innermost disk $\leq 2 {\rm AU}$, where radius of the X ray-emitting region $R_{\rm xr}$ $\approx 10 \, R_{\odot}$ is comparable to the disk thickness. In this region, the disk flare angle $\cos\theta = R_{\rm xr}/r $. The second region is the outer disk beyond $\sim 2 {\rm AU}$, where the flare angle is given by $\cos \theta = 4 \,( dH/dr - H/r ) \equiv  4 \, r\,  d \,(H/r) /d r $ (Lesniak \& Desch 2011). We make the following assumption that scale height $H$ varies with $r$ as $H \approx H_0 \,(r/1\,{\rm AU})^{1.25} $ with $H_0 \approx 0.02$ at 1 AU, to derive the following equations for $\cos \theta $, whose solutions match at $r \approx$ 2 AU:
\begin{align}
 \cos \theta = 0.047 \left ( \frac{r}{1\,\rm AU} \right)^{-1} , r \leq 2{\rm AU} \\
 \cos \theta = 0.02 \left ( \frac{r}{1\,\rm AU} \right)^{0.25},  r \geq 2{\rm AU}.
\end{align}
We implement a steady state ionization-recombination equilibrium with gas-grain chemistry by considering ionization by X rays and cosmic rays, and recombination of ions and electrons in the gas phase and on dust grains, with the following equations:
\begin{equation}
\frac{dn_e}{dt} = \zeta n_{H_2} - n_e n_{\rm gr} \pi a^2_{\rm gr} C_e S_e \tilde{J_e}  - \beta_g n_e n_i
\end{equation}
\begin{equation}
\frac{dn_i}{dt} = \zeta n_{H_2} - n_i n_{\rm gr} \pi a^2_{\rm gr} C_i S_i \tilde{J_i}  - \beta_g n_e n_i
\end{equation}
Ionization and recombination are both assumed to quickly establish equilibrium, and hence rates of change in electron and ion density ($dn_e/dt$ and $dn_i/dt$) on the left hand side of the Equations 18 and 19 are assumed to be 0. Here, $\zeta$ is the sum of the ionization rates due to all ionizing sources, $n_{\rm gr}$ is the number density of dust grains and $a_{\rm gr}$ is the size of the dust grain assumed to be 1$\,\mu{\rm m}$. $C_k$ and $S_k$ are the thermal velocity and sticking coefficient of species $k$ respectively. $\tilde{J_k}$ is the collision cross section of $k$, taken from Draine \& Sutin (1987) who consider the effects of grain charging on the probability of collisions of ions and electrons on dust grains. $\beta_g$ is the gas phase recombination coefficient. In each zone, overall charge neutrality is assumed to be quickly attained.

When dust is absent, the above equation reduces to a simple ion-balance equation with only gas-phase chemistry:
\begin{equation}
n_i(z)= \left [ \frac {\left( \zeta_{\rm xr} + \zeta_{\rm cr} + \zeta_{\rm rad} \right) n_{H_2} } {\beta_g} \right]^{1/2}  {\rm cm}^{-3},
\end{equation}
where we have also included a small $\zeta_{\rm rad} = 7 \times 10^{-19}\, {\rm s}^{-1}$ to account for ionization due to radioactive decay of $^{26}{\rm Al}$, consistent with Umebayashi \& Nakano (2009) . We assume that the abundances of short-lived radionuclides like $^{26}{\rm Al}$ are uniform across the disk, which may not be true, depending on the spatial distribution of the sources of radionuclides, such as one or more nearby supernovae or AGB stars, and their time of injection, or spallation reactions within the protoplanetary disk (Davis \& McKeegan 2014). Ionization by radionuclide decay is in any case a minor contribution.

We calculate the number density of hydrogen molecules as, $n_{H_2}(z) = \rho(z)/1.4\, m_{H_2}$. 
In each zone, our dust chemistry routine solves Equations 18 and 19 iteratively for the equilibrium abundances of ions and electrons, and calculates the charge on dust grains, at each $r$ and $z$ in the disk. The grain abundance $n_{\rm gr}$ at each $r$ and $z$ is decided by the gas-to-dust mass (g/d) ratio assumed. This computationally intensive step of directly calculating the equilibrium abundances within the disk evolution code itself motivated our choice for picking a lower number of radial (60) and 25 height zones.

We use a range of values for the gas-to-dust mass ratio (g/d): [100,1000,10000] in order to explore the evolution of the disk in different stages of grain growth, centering on a value of 1000. As the extent of recombination depends primarily on the grain surface available, changing (g/d) while keeping grain size $a_{\rm gr}$ constant is equivalent to changing $a_{\rm gr}$ with a constant (g/d), as both $a_{\rm gr}$ and (g/d) affect the total available dust grain surface area.  We also assume a range of values for $\beta_g$ to account for disk chemistry that focuses on two different ionic species : i) molecular ion chemistry, where the fast recombination reaction of HCO$^+$ with electrons is considered ($\beta_g$ = $1 \times 10^{-6} \, {\rm cm^{3}\, {\rm s}^{-1}} $); ii) metal ion chemistry, where it is assumed that all molecular ions have transferred their charge to metal ions in comparatively fast charge-transfer reactions, following which these metal ions recombine very slowly with electrons ($\beta_g$ = $3 \times 10^{-11}$/$\,T^{1/2})\,{\rm cm}^{3}\, {\rm s}^{-1}$ ; here the slow metal-electron recombination reaction dominates the rate; and iii) a simple reduced chemistry network with both species (molecular and atomic ions) that aims to replace both populations with a single species having an intermediate effective $\beta_{g,{\rm eff}}$ coefficient that will serve to generate electron densities similar to those attained when both species are present. For this paper, we assume this intermediate effective $\beta_g$ is $10^{-8} \,{\rm cm^{3}\, {\rm s}^{-1}} $, taken as the approximate mean between 10$^{-6}\, {\rm cm^{3}\, {\rm s}^{-1}}$ and 10$^{-11}\,{\rm cm^{3}\, {\rm s}^{-1}} $ for molecular-ion and metal-ion dominated chemistry, respectively. 

We also explore the effects of different ionization rates on disk structure, as well as the effect of exclusion of cosmic rays due to stellar winds (Cleeves et al.\ 2014, 2015) by turning off cosmic rays altogether.

\subsubsection{$\alpha$ derived from MRI viscosity}

To calculate $\alpha(r)$, we first calculate the ion density $\rho_i =  n_i(z)\, m_i $, in each zone of the disk, where $m_i = 23\, m_H$. Thereafter, we incorporate the numerical results from Bai \& Stone (2011) who consider the effect of non-ideal magnetohydrodynamical (MHD) phenomena in the evolution of magnetorotational instabilities in a protoplanetary disk. Non-ideal MHD effects are especially important in protoplanetary disks which are only partially ionized by cosmic ray and stellar X rays. In 3D shearing box simulations, Bai \& Stone (2011) incorporated the effect of ambipolar diffusion via the parameter $Am$, that represents the collision frequency of ions and neutral particles in one orbital period:
\begin{equation}
Am = \frac{\gamma \rho_i} {\Omega}. 
\end{equation}
Here, $\gamma=3.5 \times 10^{13}\; {\rm cm}^{3}\,{\rm s}^{-1}\, {\rm g}^{-1} $ is the drag coefficient for ion-neutral collisions (Blaes \& Balbus 1994; Draine, Roberge \& Dalgano 1983)    
From their numerical simulations, Bai \& Stone (2011) find that when turbulence is in saturation in the disk, a strong correlation is found between the turbulence stress factor $\alpha$ (from Shakura \& Sunyaev 1973) and the ratio of the gas to magnetic pressure $\beta$:
\begin{equation}
\alpha= \frac{1}{2\,\beta_{\rm min}} \,, 
\end{equation}
where $\beta_{\rm min}$ is the minimum bound of $\beta$ below which the magnetic field is too strong to be destabilized by the MRI. From the results of all their simulations, they find a fitting function correlating $\beta_{\rm min}$ and $Am$:
\begin{equation}
\beta_{\rm min}(Am) = \left [ \left( \frac{50}{Am^{1.2} }\right)^2 + \left(\frac{8}{Am^{0.3} } + 1 \right)^2 \right] ^{1/2} 
\end{equation}
Using the above Equations 21 - 23, we calculate a local $\alpha$ at all locations in the disk. Then we compute a vertically integrated and mass-weighted value $\langle\alpha(r)\rangle$ across the height of the disk, that is a function of $r$.
We impose a floor of $1 \times 10^{-5}$ on $\alpha$ in the inner disk, without which the interior of the disk evolves so slowly that it affects the numerical stability of the code.

\subsection{Photoevaporation}

We implement external photoevaporation due to FUV radiation from nearby massive stars, using the equations for photoevaporative mass loss rates for sub-critical disks (i.e., when disk radius $r_d \ll r_g$, the gravitational radius) from Adams et al.\ (2004) as follows:
\begin{equation}
\dot{M_{\rm pe}} = C_0 \,N_C\, \langle \mu \rangle \, c_s \,r_g \left(\frac{r_g}{r_d}\right) \exp \left( -\frac{r_g}{2r_d} \right) ,
\end{equation}
where we assume $C_0 = 4$, $r_d$ is the disk edge, $N_C=1.25 \times 10^{21} {\rm cm}^{-2}$ is the critical column density for the attenuation of FUV, and $r_g$ is the radius at which gas molecules are sufficiently thermally excited to be able to escape the gravitational potential of the star, given as :
\begin{equation}
r_g = \frac{G M_{\ast} \langle \mu \rangle}{kT_{\rm FUV}}    \;\;\; {\rm AU} .
\end{equation}
Here $c_s$ = $(k \,T_{\rm FUV}\,/\langle\mu\rangle)^{1/2}$ and $\mu= 1.25\, m_H$, where $m_H$ is the mass of a hydrogen atom.

The FUV flux is usually expressed as $G_0$, normalized to the Habing field, where 1 Habing field $\,= \,1.6 \times 10^{-3}\, {\rm ergs\,cm}^{-2}\,{\rm s}^{-1}$. The average flux of the interstellar FUV radiation field is equivalent to $G_0$= 1.7 Habings.
It is not simple to estimate the temperature of the photoevaporating disk atmosphere due to FUV radiation. From the temperature vs. optical depth profiles in Figure 2 of Adams et al. (2004), temperature is seen to be extremely sensitive to $G_0$. We estimate a $T_{\rm FUV}$ dependence with $G_0$ by assuming an average number density $\sim 10^{-4}$, as follows:
\begin{equation}
T_{\rm FUV}= 250 \left(\frac{G_0}{3000} \right)^{0.5} {\rm K} .
\end{equation}
In the treatment of photoevaporation adopted in this work, we only include the $\dot{M_{\rm pe}}$ from the disk edge and do not include any $\dot{M_{\rm pe}}$ from the top and bottom surfaces of the disk such as that given in Appendix A in Adams et al. (2004). Using Equations A7 and A8, we find that $>$75\% of mass is lost from the outer edge. However, we also find that Equation A8 makes inconsistent assumptions about the geometry of the flow that likely overestimates the mass lost from the surface of the disk.
We have used a range of $G_0$ in this study of [300, 1000, 3000]. This is mainly motivated from the results of Adams et al. (2006, Figure 9) where they find that the median flux experienced by a cluster star is $\sim$ 1000. 300 and 3000 represent particular values below and above this median used by Adams et al.\ (2004).

\section{Results}

In this section, we describe the results obtained from our numerical simulations of disk evolution where we explore the effect of external photoevaporation (due to FUV radiation from nearby massive star), non-uniform $\alpha$ due to the MRI and dust chemistry.

Our numerical simulations can be best categorized as two sets of disk evolution simulations: one performed with the usual uniform $\alpha$ standardization, and one where we include an MRI derived viscosity treatment from which we obtain a non-uniform variable $\alpha$ as a function of time and disk radius $r$. We have also performed several sets of simulations to assess the effects of important parameters in our simulations that are likely to have significant impact on disk structure or are known to have a range of possible values via observations. For the uniform $\alpha$ cases, we have explored the effects of different values of $\alpha$ and different radiation environments (via the parameter $G_{0}$). For variable $\alpha$ cases, we have tested the effects of different values of $G_{0}$, gas-to-dust (g/d) ratio, gas phase recombination coefficient $\beta$ (to account for atomic or molecular-ion disk chemistry), change in ionization rates due to different stellar X-ray luminosities $L_{\rm xr}$ and cosmic radiation. We vary many of these parameters by an order of magnitude above and below a canonical value. Table 1 summarizes all the simulations carried out in this investigation. We quantify the effects of variation of these parameters by looking at how they change the following disk properties related to its structure: i) mass $M_d$ of the disk; ii) slope $\langle p\rangle$ of the surface density $\Sigma(r,t)$ profile; iii) disk size or outer radius $r_d$ ; and iv) transition radius $r_T$ (the radius at which the net mass flow in the disk changes its direction from inward to outward (described in detail below).

\subsection{Uniform $\alpha$}

The following section describes the simulations performed with uniform $\alpha$, i.e., runs 1, 2, 3 and 4 [See Table 1].
\subsubsection{Canonical Case for Uniform $\alpha$}

We present runs 1 and 2 as the canonical case for disk evolution with uniform $\alpha$. Run 1 is a simple uniform $\alpha$-disk that viscously expands with time. In run 2, this disk is subjected to external photoevaporation with an FUV flux of $G_0=1000$ (assuming the median value from Fig.\ 9 in Adams et al.\ 2006). In both runs we adopt $\alpha = 10^{-3}$ as a typical value of $\alpha$ considered averaged throughout $r$. This choice of $\alpha$ is consistent with the range of $\alpha$ inferred from observations of resolved disks from Andrews et al.\ (2009, 2010). 

Fig.\ 1 and Fig.\ 2 show the typical surface density profile $\Sigma(r,t)$ for a uniform $\alpha$-disk undergoing viscous evolution without photoevaporation ($G_0=1$), and with photoevaporation ($G_0=1000$), respectively. While a non-photoevaporated disk viscously expands with time (Fig.\ 1) and loses mass mainly via accretion onto the star, a photoevaporated disk loses mass to both accretion onto the central star, as well as photoevaporation via the outer edge of the disk over 10 Myr, as seen in Fig.\ 2 and Fig.\ 3. For the photoevaporated disk, the rates for mass loss due to accretion and photoevaporation are both similar, i.e., $\dot{M_{\rm acc}} \approx \dot{M_{\rm pe}} \sim 10^{-8}\, \rm{M_{\odot} \,yr^{-1}}$. The dip in each curve in Fig.\ 3 represents the transition radius, $r_T$, i.e., where the directionality of the net mass flow changes from inward towards the central star to outward. 

In order to monitor the average slope of the $\Sigma$ profile of the disk, for each simulation, we also plot  $\langle p \rangle$ = $d\,({\rm log} \,\Sigma )/ d({\rm log} \,r)$ with time, in Fig.\ 4, where $\langle p \rangle$ is spatially averaged over the giant planet formation region, i.e., 5-30 AU (discussed in Desch 2007). Since we use $\langle p \rangle$ across the region 5-30 AU, curves for $\langle p \rangle$ with time for all simulations are plotted till the disk is truncated to 30 AU. 

A feature that stands out in this set of simulations is that the profile of the disk and its slope remains uniformly preserved throughout its 10 Myr of evolution, as is seen in Figs.\ 2 and 4.  The slope $\langle p \rangle$ is almost constant ($\sim$ 1.6) throughout the simulation duration, although a slight increase is noted in the last few Myr of simulation. The profile of the non-photoevaporated disk, in contrast, is seen to flatten towards $\sim$ 1 with time, consistent with theoretical predictions. Fig.\ 4 also shows the change in disk mass with time for both the non-photoevaporated and photoevaporated cases. As expected, the mass of a photoevaporated disk is considerably lower after 10 Myr, as compared to a non-photoevaporated disk. Fig.\ 4 additionally shows how the outer radius $r_d$ and transition radius $r_T$ varies as evolution proceeds. A non-photoevaporated disk viscously expands with time (as seen in Fig.\ 1; not shown in Fig.\ 4 as our simulations are only performed to a radius of 100 AU). A photoevaporated disk on the other hand, continually shrinks in size with time due to continuous removal of mass from the outer disk edge by photoevaporation. $r_T$ varies distinctly in both cases, by moving outward with time in a non-photoevaporated viscously spreading disk (see Equation 23 in Hartmann et al.\ 1998) and moving inward with time in a photoevaporated disk. This leads us to a picture where more and more mass moves outwards in a photoevaporated disk, as the disk itself shrinks in size. 

\subsubsection{Parameter Study: Effect of $\alpha$}

Runs 2, 3 and 4 explore the effect of variation of the parameter $\alpha$ in photoevaporated ($G_0=1000$) uniform $\alpha$ disks. Fig. 5 shows how different disk properties vary with time with different values of globally-averaged $\alpha$ in the disk. As expected, higher the value of $\alpha$, more rapid is the disk evolution so much so that disks with $\alpha$ $\sim$ $10^{-2}$ dissipate within $\sim$ 4 Myr. Such a disk loses more than 95\% of its mass within 2 Myr. This rapid evolution and movement of most of its mass is also indicated by the rapid change in the slope $\langle p \rangle$ of the 5-30 AU region. On the contrary, disks with $\alpha=10^{-4}$ evolve so slowly that they lose $<$ 40\% of their mass in 10 Myr. $\langle p \rangle$, $r_T$ and $r_d$ in these disks remain more or less constant. Disk simulations with $\alpha=10^{-3}$ show an intermediate behavior between the two extremes, by retaining $\sim$ 10\% of its mass after 10 Myr with a slowly increasing slope $\langle p \rangle$ $\sim$ 1.5, shrinking to a final size of $\sim$  60 AU after 10 Myr. $r_T$ shows a dual behavior as it initially moves outward over the first 5 Myr, and thereafter moves inwards with time.

Nevertheless, a uniform value for $\alpha$ is not realistic, and we hereafter present simulations where we look at the effect of a radially and temporally varying $\alpha$ on disk structure.

\subsection{Non-uniform $\alpha$}

The following section describes in detail simulations performed with computed $\alpha$, i.e., runs 5-16 [Table 1].

\subsubsection{Canonical Case for Non-uniform $\alpha$: A dust-free disk}

Runs 5 and 6 represent the dust-free simulations for varying $\alpha$ shown in Figs.\ 6-11 taking $L_{\rm{xr}}=1$ $\times$ $10^{29} \,{\rm ergs}\,{\rm s^{-1}}$,  $\beta_g=10^{-6}\,{\rm cm}^{3}\, {\rm s}^{-1}$, with $G_0=1$ for the non-photoevaporated case, and $G_0=1000$ for the photoevaporated case. The photoevaporated disk was evolved for $\sim 7.5$ Myr, after which the simulation was terminated when $r_d$  approached $<  5 {\rm\,AU}$.

Figs.\ 6 and 8 shows the variation of a vertically-averaged mass-weighted $\langle \alpha \rangle$ derived from the MRI (as described in Section 2) with $r$ at 3 different times in disk evolution for $G_0=1$ (Fig.\ 6) and $G_0=1000$ (Fig.\ 8). Initially at $t=0$,  $\langle \alpha \rangle $ varies considerably across the disk, from $\sim$ few $\times$ $10^{-3}$ in the inner disk, to $\sim 10^{-2} $ in the outer disk, in both the non-photoevaporated and photoevaporated cases. This is due to the difference in the ionization fraction between the poorly-ionized dense self-shadowing inner disk and the highly ionized tenuous outer disk. At $t_{mid}= 4-5 $ Myr, in both cases, the inner disk also attains a higher value of $\alpha$ as much of the inner disk mass is cleared out due to accretion. Thereafter, in the last few Myr, $\alpha$ begins to settles to a constant value of $10^{-2}$ throughout the disk. High values of $\alpha$ in the outer disk result in increased turbulent mixing and therefore rapid mass movement in the outer disk. Fig.\ 7 shows $\Sigma(r,t)$ for a non-photoevaporated disk, in which the disk shows comparatively quick dissipation even without photoevaporation. With photoevaporation (Fig.\ 9) however, very rapid dissipation of the disk truncates the disk to $<$ 5 AU within 7.5 Myr. This occurs because the high value of $\alpha$ in the outer disk makes it easier for external photoevaporation to remove more mass from the outer edge causing quick disk dispersal. Fig.\ 10 shows $\dot{M}$ profiles for the photoevaporated disk. The inner disk with lower values of $\alpha$ allows little movement of mass, and hence less mass flow results in the inner regions. $\Sigma(r,t)$ from Fig.\ 9 and $\langle p \rangle$ vs. time plot from Fig.\ 11 show that the 5-30 AU slope of the disk profile is maintained at $\langle p \rangle$ $\sim 1.75$ for upto 1.5 Myr, after which it steepens sharply to $\langle$p$\rangle \sim 3.0$ due to increase in mass loss by photoevaporation. Thereafter, as the disk shrinks in, the inward mass loss due to accretion becomes greater than the photoevaporative mass loss rate, ultimately flattening the slope profile towards the end of the simulation. However, the overall disk structure (Fig.\ 9) is seen to be mostly maintained through the rapid disk dissipation. Fig.\ 11 also shows the $M_d$ vs. time and $r_d$ and $r_T$ vs. time. $r_T$ is seen to move inwards during the first 5 Myr.

\subsubsection{Canonical Case for Non-uniform $\alpha$: Gas+Dust disk}

Runs 7 and 8 (Figs.\ 12-17) show disk evolution simulations for varying $\alpha$ where dust has been included in the disk. We adopt a uniform grain size of $a_{gr}$ = 1$\mu$m and a gas-to-dust (g/d) ratio of 1000. Our choice for a larger value of (g/d) than standard is motivated by the fact that it was comparatively difficult to obtain any significant disk evolution with the standard g/d of 100. While this choice of (g/d) could be assumed to be a slightly advanced stage in grain growth, we note that the standard g/d=100 usually assumed in protoplanetary disks is itself an assumption. We also assume a $G_0=1000$ as the typical FUV flux incident on the disk, similar to the previous sets of simulations. A stellar X-ray luminosity of $10^{29}$ ergs s$^{-1}$ and an effective $\beta_g$ of $10^{-8} \,{\rm cm}^{3}\, {\rm s}^{-1}$ are assumed, as discussed in Section 2. 

It was required to impose a floor value on $\alpha$ to assist disk evolution in the inner disk, as the addition of dust made the evolution of the disk interior very slow (Figs.\ 12 and 14). The inner disk due to its high density is weakly ionized. Stellar X-rays and cosmic radiation are able to penetrate the cloud only where it is optically thin. The presence of dust makes this already scarce availability of charges worse by absorbing them and hence maintaining a very small ionization fraction of ions and electrons in the disk interior. Rates of infall onto the star plummet to $\sim$ few $\times$ $10^{-10}$ $\rm{M_{\odot}}^{-1}$ (Fig.\ 16). However, the outer disk being optically thin is sufficiently ionized by both cosmic radiation and oblique stellar X rays which drives rapid mass flow. Figs.\ 15 and 17 show this more clearly, as the presence of dust chokes inner disk evolution such that a significant fraction of mass in the disk is only redistributed towards the inner disk. This causes $\alpha$ to rise dramatically in a sharp transition from $10^{-4}$ to 0.1 in the 3 AU - 20 AU region; which moves inward with time (Fig.\ 14). As more and more mass falls onto the star, the inner disk becomes less dense enough to be sufficiently ionized. Movement of mass picks up and $\dot{M_{\rm acc}}$ go up by an order of magnitude in the inner disk. From Fig.\ 17, we see $r_T$ move inward from around 7 AU at 1 Myr to 3 AU, as the disk radius $r_D$ reduces to $\sim$ 50 AU. It is interesting to note how not only does the mass of the disk drop almost linearly with time, it keeps up this linearity with increase in $G_0=1000$ as well. 
From the above simulations, we see that unlike dust-free simulations, a dusty disk does not lose much mass with time (Fig.\ 17).

\subsection{Effect of each parameter in a [gas+dust] disk with computed $\alpha$}

\subsubsection{Effect of $G_0$}

In runs 8, 9 and 10, we vary the flux of external FUV radiation illuminating the disk through $G_0$ = 300, 1000 and 3000. These results are plotted in Fig.\ 18 to show how various disk structure properties vary with time.
Higher $G_0$ causes more mass loss in the outer disk. The high values of $\alpha$ between 0.001 to 0.1 due to high ionization fractions in the outer disk likely facilitates this rapid mass movement, and eases outward mass loss due to photoevaporation.
As noted before, the linearly decreasing trend in the disk mass with time is kept up with an intermediate value of $G_0=300$, as well as $G_0=3000$ as well. Different values of $G_0$ typically show very steep initial slopes, flattening almost similarly with time. While higher $G_0$ truncates the disk to a smaller $r_d$, $r_T$ does not show any such trend with increasing $G_0$. The $r_T$ for $G_0$ of 1000 and 3000 are mainly similar throughout the simulation. 

\subsubsection{Effect of gas-to-dust (g/d) ratio}

Runs 8, 11 and 12 show the effect of varying the gas-to-dust mass (g/d) ratio (Fig.\ 19). In these runs, we can see that the disk is effectively cleared within 10 Myr only when g/d = 10000. Higher g/d can be taken to be a proxy for grain growth. Therefore, it seems likely that rapid evolution of the inner disk is only possible with substantial grain growth. g/d=100 shows the other extreme case where the inner disk evolution is too slow that the mass in the outer disk is piled on top of the mass in the inner and middle disk creating an extremely steep slope. The outer disk edge $r_d$ does not change much with two orders of variation of (g/d) ratio till about halfway through the simulation. The transition radius $r_T$ decreases from $\sim$ 15 AU after 1 Myr to $\sim$ 2 AU at 10 Myr.

\subsubsection{Effect of $\beta_g$}

Runs 8, 13 and 14 show the effect of varying the gas phase recombination coefficient $\beta_g$ (Fig.\ 20) exploring the extremes of $\beta_g$ between molecular-ion dominated chemistry ($\beta_g = 10^{-6} \,{\rm cm}^{3}\, {\rm s}^{-1}$) and metal-atom dominated chemistry ($\beta_g = 10^{-11} \,{\rm cm}^{3}\, {\rm s}^{-1}$). The highest value of $\beta$ corresponding to recombination of molecular ion ${\rm HCO^+}$ causes the disk to lose about half its mass compared to the other recombination rates. Slope varies wildly throughout r for the two lower values of $\beta_g$. Disk radius $r_d$ and transition radius $r_T$ do not seem to be very sensitive to gas recombination rates. 

\subsubsection{Effect of $L_{xr}$}

Runs 8 and 15 vary the stellar X-ray luminosity by an order of magnitude, i.e. $10^{28}$ ergs s$^{-1}$ (Fig.\ 21). We find that all disk properties (mass $M_d$, radius $r_d$, transition radius $r_T$ and slope) that are tested are not sensitive to the change in X-ray luminosity.

\subsubsection{Effect of Cosmic Ray Exclusion}

In run 16 (Fig.\ 22), we removed cosmic radiation from our model to note the effect of the exclusion of cosmic rays on disk structure. While disk mass and outer radius does not seem to be sensitive to the presence or absence of cosmic radiation, the internal structure of the disk is still impacted by the absence of cosmic rays. Cosmic rays appear to be the primary source of ionization for the mid-regions of the disk (6-20 AU) that are not as optically thick as the disk interior. In the absence of cosmic rays, the disk develops very steep profiles over most of the simulation timescale (10 Myr), as the middle regions of the disk are much less ionized than in the presence of cosmic rays. The outer flared disk is still ionized by X-rays and spreads to pile up on the relatively static middle disk, steepening the profile. 

\section{Conclusions}

In this work, we have performed evolutionary simulations of protoplanetary disks subject to the influence of external photoevaporation (due to FUV radiation from a nearby massive star), and non-uniform viscosity due to the magnetorotational instability (MRI). For the latter, we have calculated the equilibrium ionization state at each radius $r$ and height $z$ of the disk with a simple gas-grain chemistry network. External photoevaporation is an efficient disk dispersal mechanism, and can dramatically alter disk evolution by steadily removing mass, and truncating the size of the disk. Rapid mass removal dictated by photoevaporation places lower limits on planet formation timescales due to disk dissipation. Half of all protostars are likely to be born in rich clusters containing at least one O star that would photoevaporate their protoplanetary disks (Lada \& Lada (2003). There are also numerous lines of evidence that suggest that Sun's disk was photoevaporated in the past. Photoevaporation is therefore important to be included in disk models in order to understand the evolution of the solar nebula. 
It is also important to note that most disk models employ a uniform $\alpha$, usually 0.01, as the coefficient of turbulent viscosity, which is not attributed to any particular physical mechanism. All considered physical processes would yield a non-uniform $\alpha$ through the radial extent of the disk. In this work, the widely-accepted magnetorotational instability (MRI) is taken to be the mechanism driving the angular momentum transport and viscosity in the disk. The operation of the MRI depends on the local ionization state of the disk and thus varies with $r$ and $z$, causing $\alpha$ to also vary with $r$ and $z$.  Using the formulations of Bai \& Stone (2011) that include non-ideal MHD with ambipolar diffusion, we find that a varying $\alpha$ profile can drive mass movement dramatically differently in the disk and can therefore significantly affect disk structure. 
\subsection{Main results}
Our simulations have explored the structure and evolution of a 0.1 ${\rm M_{\odot}}$ disk around a 1 ${\rm M_{\odot}}$ star over several Myr timescales, under the influence of a range of irradiating stellar birth environments and the inclusion of a prescription for non-uniform MRI-viscosity. The main results of our simulations incorporating all of the above effects are summarized below:
\begin{itemize}
\item We find that turbulent viscosity $\alpha$ derived from the MRI can vary over four orders of magnitude with $r$, i.e., from $< 10^{-5}$ in the inner disk to 10$^{-1}$ in the outer disk. This variation is due to the difference in the ionization fraction throughout the disk. While the dense shadowed inner disk is largely neutral, the outer tenuous disk is highly ionized by both cosmic radiation, as well as X rays that reach the flared outer disk. This variation in $\alpha$ causes mass to move very slowly in the inner disk, but simultaneously drives rapid mass movement in the outer disk. Such a variation in mass transport creates an inherently steep profile in the early disk. 
\item Photoevaporation due to $G_0 = 1000$ is able to rapidly remove mass from the outer disk edge in timescales of $\sim$ few Myr, and truncate the disk outer edge, to $\sim$ 50 AU in 10 Myr in a dusty disk. In a disk with little or no dust, it rapidly dissipates the disk down to 5 AU within $\sim$ 7.5 Myr. Over and above the steep disk profile created by non-uniform $\alpha$, photoevaporation steepens the slope in the outer disk (i.e., 5 - 30 AU) due to removal of mass from the other edge, but by not as much as due to the radially varying $\alpha$.
\item An interesting result from our simulations is that in a photoevaporated disk, the transition radius $r_T$ (i.e., the radius at which the direction of mass flow changes from inward into the star to outward) can move inward with time. This implies that external photoevaporation is able to move mass from the inner disk to the outer edge. This is unlike the case of a non-photoevaporated viscously spreading disk (e.g., LBP) where $r_T$ moves outward with time. 
\item Higher FUV fluxes (e.g., $G_0 = 3000$) remove more mass from the disk, bringing the outer radius $r_D$ as well as the transition radius $r_T$ inward, and create steeper disk profiles in the outer disk. Lower FUV fluxes ($G_0=300$) have the opposite effect: they remove less mass from the disk and cause shallower disk profiles.
\item Dust influences disk evolution by absorbing charges and drastically reducing the ion fraction in the dense disk interior. In our canonical case (g/d=1000 and $a_{\rm gr}$ = 1$\mu$m), we find that the presence of dust lowers $\alpha$, hindering inner disk evolution; infall onto the star plunges below 10$^{-9}\, \rm{M_{\odot} \,yr^{-1} } $. The effects of MRI-viscosity and external photoevaporation cause rapid dissipation of a dust-free disk within 7.5 Myr. In contrast, it is able to remove only about half of the disk mass in a dusty disk over 10 Myr. The presence of dust is thus able to create two different evolutionary pathways: a stalled evolution in the case of dusty disks, and rapid dispersal in the case of dust-free disks. 
\item It is important to note that we do not consider grain growth in our simulations (discussed later). However, employing a range of gas-to-dust ratio (g/d = 100, 1000, 10000) can be a proxy for grain growth. Our results show that grain growth must necessarily proceed efficiently until sufficient evolution of the inner disk is allowed. Without efficient grain growth, the disk undergoes a stalled evolution where there is a large scale transport of mass from the outer higher-$\alpha$ disk to the lower-$\alpha$ interior.
\item The value of $\alpha(r)$ can be affected by several factors, including how ionization processes ionize the disk across $r$ as well as models of disk chemistry used in the simulations. An order-of-magnitude variation in ${L_{\rm xr}}$ does not affect the overall disk structure in a dusty disk. Cosmic ray exclusion, however, causes steep profiles in the mid-regions of the disk that are important for giant planet formation. Changing the disk chemistry also results in wildly varying slopes of the $\Sigma$ profile with time. 
\end{itemize}
\subsection{Implications}
Our results show that the disk evolves very differently from previous disk models such as the self-similar viscous evolution models (LBP; Hartmann et. al 1998) under the combined action of both non-uniform viscosity with dust and external photoevaporation, each of which contributes toward steepening of the disk profile. Here, we highlight some of the most important implications due to the variation of the structure and evolution of the non-uniform $\alpha$-disk subject to photoevaporation.
\subsubsection{Changes in the Disk Structure and Mass Transport}
We find that the presence of dust dictates two distinct evolutionary tracks for non-uniform $\alpha$ disks subject to external photoevaporation. We describe each case separately as follows:

In the case without dust, the structure of the disk is significantly altered due to the difference in mass flow between the inner dense disk and the outer tenuous disk. In a dust-free disk, $\alpha$ is an increasing function with $r$ for several Myr, due to the differing ion fractions between the inner and outer disk, creating an initial steep surface density profile as seen in Fig.\ 8. $\alpha$ ranges over 2-3 orders of magnitude ($\sim$ few $\times$ 10$^{-4}$ in the inner disk to $\sim$ 10$^{-1}$ in the outer disk). This difference in magnitude is eventually reduced with time and the profile flattens out as the inner disk is cleared out by accretion onto the star, increasing the ion fraction in the interior and thus $\alpha$. Determining the viscous timescale $t_{\rm visc}$ across the disk can lend a quantitative insight into the timescale of this mass transport process, for which the following expression is useful:
\begin{equation}
t_{\rm visc} = \frac{r^2}{\nu} \equiv \frac{(r/H)^2}{\alpha \Omega}
\end{equation}
Assuming the disk is flaring slightly, using $\nu = \alpha \,H^2 \,\Omega$ (from the parameterization of Shakura \& Sunyaev (1973), as well as $c_s$ = $H\,\Omega$), if $\alpha$ is a constant, then $t_{\rm visc}$ $\propto$ $r^{3/2}$. When we assume a non-uniform value of $\alpha$, in a disk with little or no dust, we find $\alpha$ varies as $\alpha$ $\propto$ $r$ between 2 - 20 AU at $t = 0$. This leads to an initial value of $t_{\rm visc}$ $\propto$ $r^{0.5}$. Later at $t = 4$ Myr, as the $\alpha$ profile steepens to $\sim$ $r^{1.4}$, $t_{\rm visc}$ $\propto$ $r^{0.1}$.  
The overall steady shape of the disk profile is maintained throughout the simulation as the rates of mass loss $\dot{M}_{PE}$ and $\dot{M}_{acc}$ match each other throughout the duration of 7.5 Myr. 

In the case with dust, as seen in Fig.\ 14, the presence of dust exaggerates the already-increasing $\alpha$ profile as dust absorbs and removes charges from the inner disk. In a dusty disk, the $\alpha$ slope is steeper, i.e., $\alpha$ $\propto$  $r^{2.0}$ at $t = 0$ from 1-50 AU and later increases to $\sim$ $r^{3}$ at $t = 5$ Myr within a 2-20 AU region. This leads to $t_{\rm visc}$ $\propto$ $r^{-0.5}$ at $t = 0$, increasing to $r^{-1.5}$ at 5 Myr. The $t_{\rm visc}$ effectively seems to decrease with radius through a large portion of the outer disk. Mass is therefore transported very rapidly from the outer disk into the inner disk (within a few AU). This is a robust result as this results from the initial steep profile that arises from the non-uniform $\alpha$. This can also be seen in Fig.\ 16 where $\dot{M}_{\rm PE}$ $>$ $\dot{M}_{\rm acc}$. The disk initially loses more mass to photoevaporation than accretion, and it takes several Myr for accretion rates to catch up to photoevaporative rates, steepening the already steep surface density profile in the outer disk. 
From the results of our simulations, we find that until grain growth is efficient, inner disk evolution is stalled and the disk develops a steep profile across the planet formation region (5 - 30 AU). If the disk evolves in this manner, while Jupiter may have sufficient $\Sigma$ in its formation region to reach isolation mass, the other outer planets may be likely left with too little mass to grow, as the disk gas is likely to be dissipated out of the outer disk very quickly. 
\subsubsection{Comparison with the MMSN profile} Our results show that disks are more likely to evolve with steeper profiles than the MMSN profile (with slope $p = 1.5$). Desch (2007) had updated the MMSN profile with the positions of the giant planets in the compact configuration of the Nice Model (Tsiganis et al.\ 2005), and had found $p$ $\sim$ 2.2. Desch (2007) had attributed this steeper slope to be due to mass removal by external photoevaporation. In this study, we investigated the evolution of the protoplanetary disks subject to photoevaporation, and a non-uniform MRI viscosity. We find that while variable $\alpha$ steepens the disk dramatically, external photoevaporation also steepens it but by not as much. The presence of dust also significantly steepens disk structure, the extent of which is uncertain as grain growth has not been included in this study.
It would be useful to compare our model disk profiles to MMEN surface density profiles derived from Kepler data, but these are pertinent only to the innermost 0.5 AU of the disk, where our models are potentially uncertain due to the assumed inner boundary condition. Nevertheless, we find in our models without dust, the inner disks (0.2 to 5 AU) evolve to a state similar to the MMEN profile of Chiang \& Laughlin (2013) with $\Sigma$ profile slope p $\sim$ 1.6 in the first 3 Myr (until the disk dissipates). In our models with dust, the innermost regions (0.2 to 5 AU) have steeper slopes p $\sim$ 2.0 - 2.2, although not as steep as in their outer regions.

\subsubsection{Planetary Growth Timescales}
In order to determine how long it takes for planet cores to grow within the surface density profiles predicted for photoevaporated non-uniform $\alpha$ dusty disks, we use the planetary growth model employed by Desch (2007) [see equations 30-34 in Desch (2007)] that implements the growth rate equations of Ida \& Makimo (1993). The eccentricity of the planetesimals is derived assuming an equilibrium between the effect of gas drag and gravitational stirring of the planet cores as given in Kokubo \& Ida (2002), and the gas drag evaluated from the Reynolds number (Re) using the prescriptions from Weidenschilling (1977a). A uniform initial size of planetesimals is assumed to be 100m, similar to Desch (2007). Desch (2007) had considered a uniform non-varying surface density as well as a viscously evolving disk to calculate the growth timescales. We improve this model by taking a self-consistently decreasing solid surface density accompanying core growth with time.

Assuming that each core only accretes planetesimals from its own feeding region, we obtain the growth profiles as shown in Fig.\ 23 for our canonical photoevaporated dusty disk case. 

From the growth profiles, we note that the cores of Jupiter and Saturn grow until 0.5 Myr, while Neptune and Uranus take up to 2 Myr to accrete all the planetesimals in their feeding regions. We also note that while photoevaporation does not affect the growth of Jupiter's core, which rapidly accretes $\sim$ 90 $\rm{M_{\oplus}}$ in 0.5 Myr due to higher local $\Sigma$, higher FUV fluxes are able to significantly stunt the growth of Saturn's core. FUV fluxes corresponding to $G_0$ $>$ 1000 may not allow Saturn to accrete sufficient solid mass matching the predicted present-day value of 9 - 22 $\rm{M_{\oplus}}$ (Desch 2007). Neptune and Uranus grow negligibly even without photoevaporation as they are not able to accrete mass quickly enough before the disk dissipates. More rapid growth timescales or migration of large planetesimals into the outer disk may be needed to explain their core growth. However, our models do not include grain growth, which could potentially make the $\Sigma$ profile more shallower allowing for more mass transported to the feeding regions of the outer ice giants, potentially leading to more core growth.

\subsubsection{Radial Volatile Transport}
Lastly, we also argue that photoevaporation also dramatically affects radial transport of volatiles. Takeuchi \& Lin (2002) have argued that if the sum of the slope $p$ of the surface density profile and the slope $q$ of the temperature profile (Equation 8) is $>$ 2, then the volatiles are transported radially outward in the nebula. In our work, we assume a typical temperature profile with $q\,=\,0.5$, and we find steep surface density profiles with slope $p$ $>$ 2. Thus, in our disks, the sum of the slopes $p + q > 2$, supporting outward volatile transport. From our simulations, we predict that photoevaporation is able to remove volatiles (such as H$\rm{_2}$O) efficiently through the disk outer edge even from as far in as the inner disk. This is a result from our simulations that in a photoevaporated disk, the transition radius $r_T$ can move inward with time as opposed to a non-photoevaporated viscously spreading disk, where $r_T$ increases with time (Hartmann et al.\ 1998) as $r_T \propto T^{0.5}$. This results in more and more mass being removed from the inner disk, the region of terrestrial planet formation. Indeed, $r_T$ in some disks go as far inward as $\sim$ 3 AU. Significant loss of volatiles from the inner disk material can severely affect the potential for future habitability of planets that form in the volatile-depleted inner disk.
\subsection{Future Work}
An important caveat of our models is that we do not yet include grain growth. In the absence of grain growth, dust efficiently stagnates inner disk evolution. Accretion is very slow and mass transported from the highly ionized outer disk just accumulates in the middle and inner disk. We predict that with grain growth, the inner disk will be able to accrete onto the star after grain growth proceeds efficiently in 1-2 Myr, and increase accretion rates such that disk evolution is quickened. This way, the steep profile erected by the initial stagnation of the disk will gradually flatten with time. Such a disk may then have enough mass and time for the growth of the four giant planets across 5-30 AU. However, it is also likely to be dissipated quickly with time (as seen in g/d=10000 case in Fig.\ 19, where a higher g/d can be considered as a more advanced stage of grain growth). 

\subsection{Summary}
In this work, we have performed simulations of protoplanetary disk evolution where we have included the effects of i) external photoevaporation ii) MRI-derived non-uniform viscosity, and iii) a simple gas-grain chemical network to calculate ionization equilibrium state in the disk. From our simulations, we argue that it is important to consider both external photoevaporation and non-uniform viscosity in disk models as each contributes strongly in altering the disk profile in a unique manner.  Models incorporating external photoevaporation and a realistic prescription of viscosity and angular momentum transport may bring us closer to the behavior and evolution of the physical processes that transpired in the solar nebula that shaped the structure of the Sun's protoplanetary disk and determined the composition of the terrestrial and the giant planets. 

\section*{Acknowledgements}
This work was supported by grants from the NASA Astrobiology Institute, Nexus for Exoplanet System Science (NExSS) and Keck Institute of Space Studies. We thank Neil Turner for helpful discussions. The results reported herein benefitted from collaborations and/or information exchange within NASA's Nexus for Exoplanet System Science (NExSS) research coordination network sponsored by NASA's Science Mission Directorate.

\begin{deluxetable}{ccccccccc}
\tabletypesize{\scriptsize}
\tablecaption{Table of Simulations}
\label{table:sims}
\tablewidth{0pt}
\tablehead{
\colhead{Run} &
\colhead{$G_{0}$} &
\colhead{$\alpha$} &
\colhead{g/d} &
\colhead{$\beta_g$} &
\colhead{$\L_{xr}$} &
\colhead{CR Exc}&
\colhead{Comment}& 
\colhead{Figures}}

\startdata
$1$ & 1 &  $10^{-3}$ & N/A & N/A & N/A & N/A & Assigned $\alpha$ without PE & Fig. 1\\
$2$ & 1000 & $10^{-3}$ & N/A & N/A & N/A & N/A & Assigned $\alpha$ with PE & Figs. 2 - 4\\ 
\tableline
$3$ & 1000 &  $10^{-4}$ & N/A & N/A & N/A & N/A &  Effect of assigned uniform $\alpha$ & Fig. 5\\
$(2)$ & 1000  & $10^{-3}$ &N/A & N/A & N/A & N/A  &\\
$4$ & 1000 & $10^{-2}$ &N/A & N/A & N/A & N/A &\\
\tableline
$5$ & 1 & N/A & N/A & $10^{-8}$ & $10^{29}$ & No & MRI $\alpha$ without PE (DUST-FREE) & Figs. 6, 7\\
$6$ & 1000 & N/A & N/A & $10^{-8}$ & $10^{29}$ & No & MRI $\alpha$ with PE (DUST-FREE) & Figs. 8 - 11\\
\tableline
$7$ & 1 & N/A & 1000 & $10^{-8}$ & $10^{29}$ & No & MRI $\alpha$ without PE + DUST & Figs. 12, 13 \\
$8$ & 1000 & N/A & 1000 & $10^{-8}$ & $10^{29}$ & No & MRI $\alpha$ with PE + DUST & Figs. 14 - 17\\
\tableline
$9$ & 300 & N/A & 1000 & $10^{-8}$ & $10^{29}$ & No & (MRI $\alpha$ + DUST + PE) Effect of $G_0$ & Fig. 18\\
$(8)$ & 1000 & N/A & 1000 & $10^{-8}$ & $10^{29}$ & No & \\
$10$ & 3000 & N/A & 1000 & $10^{-8}$ & $10^{29}$ & No &\\
\tableline
$11$ & 1000 & N/A & 100 & $10^{-8}$ & $10^{29}$ & No & (MRI $\alpha$ + DUST + PE) Effect of (g/d) & Fig. 19\\
$(8)$ & 1000 & N/A & 1000 & $10^{-8}$ & $10^{29}$ & No & \\
$12$ & 1000 & N/A & 10000 & $10^{-8}$ & $10^{29}$ & No & \\
\tableline
$13$ & 1000 & N/A & 1000 & $10^{-6}$ & $10^{29}$ & No & (MRI $\alpha$ + DUST + PE) Effect of $\beta_g$ & Fig. 20\\
$(8)$ & 1000 & N/A & 1000 & $10^{-8}$ & $10^{29}$ & No & \\
$14$ & 1000 & N/A & 1000 & $3\times10^{-11}/T^{0.5}$ & $10^{29}$ & No & \\
\tableline
$15$ & 1000 & N/A & 1000 & $10^{-8}$ & $10^{28}$ & No & (MRI $\alpha$ + DUST + PE) Effect of $L_{\rm xr}$ & Fig. 21\\
$(8)$ & 1000 & N/A & 1000 & $10^{-8}$ & $10^{29}$ & No & \\
\tableline
$16$ & 1000 & N/A & 1000 & $10^{-8}$ & $10^{29}$ & Yes & (MRI $\alpha$ + DUST + PE) CR Exclusion & Fig. 22\\
$(8)$ & 1000 & N/A & 1000 & $10^{-8}$ & $10^{29}$ & No & \\
\tableline
\enddata

\end{deluxetable}

\begin{figure}
\epsscale{0.85}
\end{figure}
\clearpage
\begin{figure}
\plotone{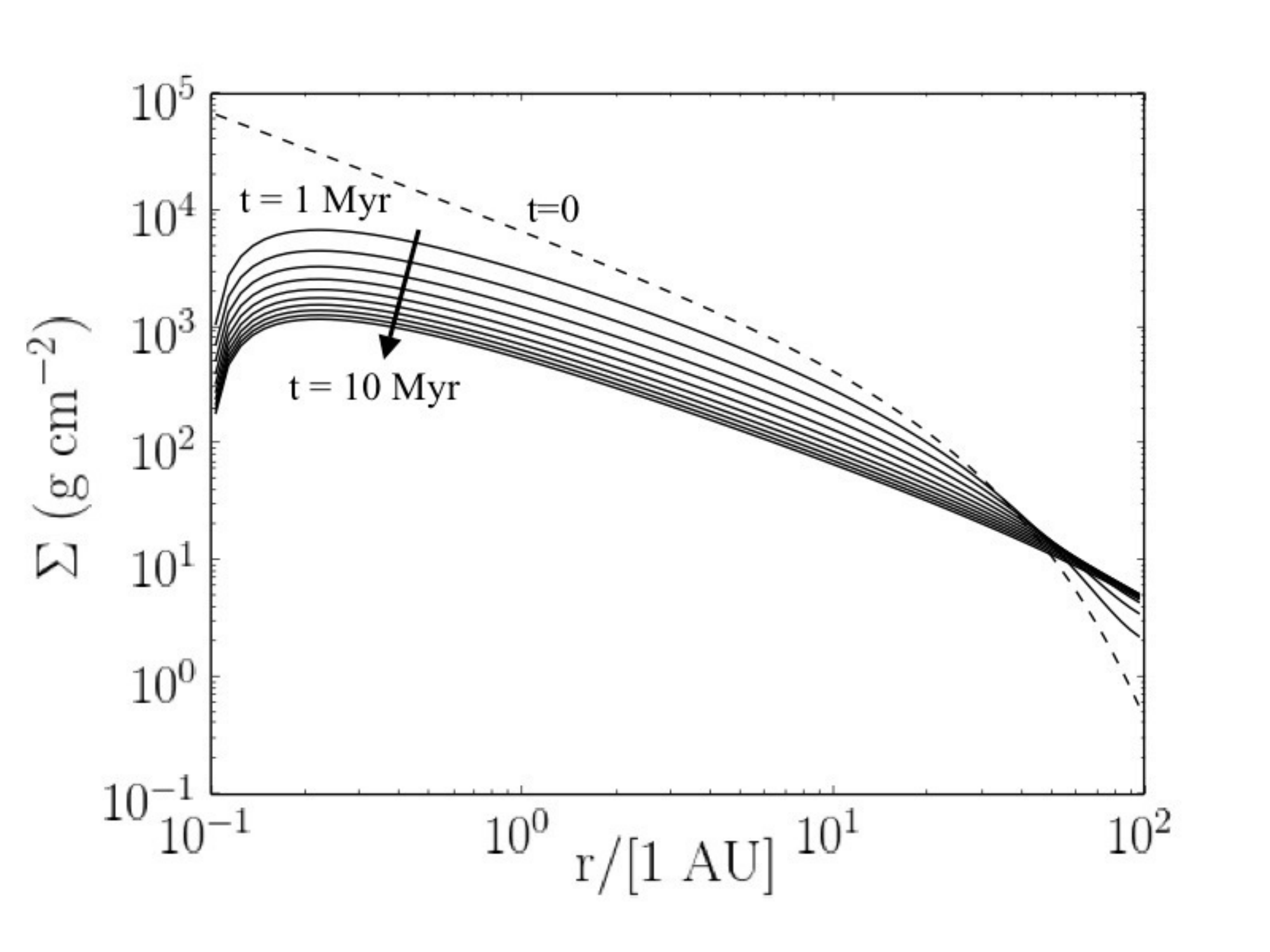}
\epsscale{0.85}
\caption{Surface density profiles $\Sigma (r,t)$ of our canonical uniform $\alpha$ case without photoevaporation. $\alpha$ is assumed to be 0.001, and $G_0 = 1$ in this run. Each curve shows the $\Sigma$ profile at times: 0 Myr (dashed), 1 Myr, 2 Myr, .. , 10 Myr. Note that the non-photoevaporated disk viscously expands with time.}
\end{figure}
\clearpage
\begin{figure}
\plotone{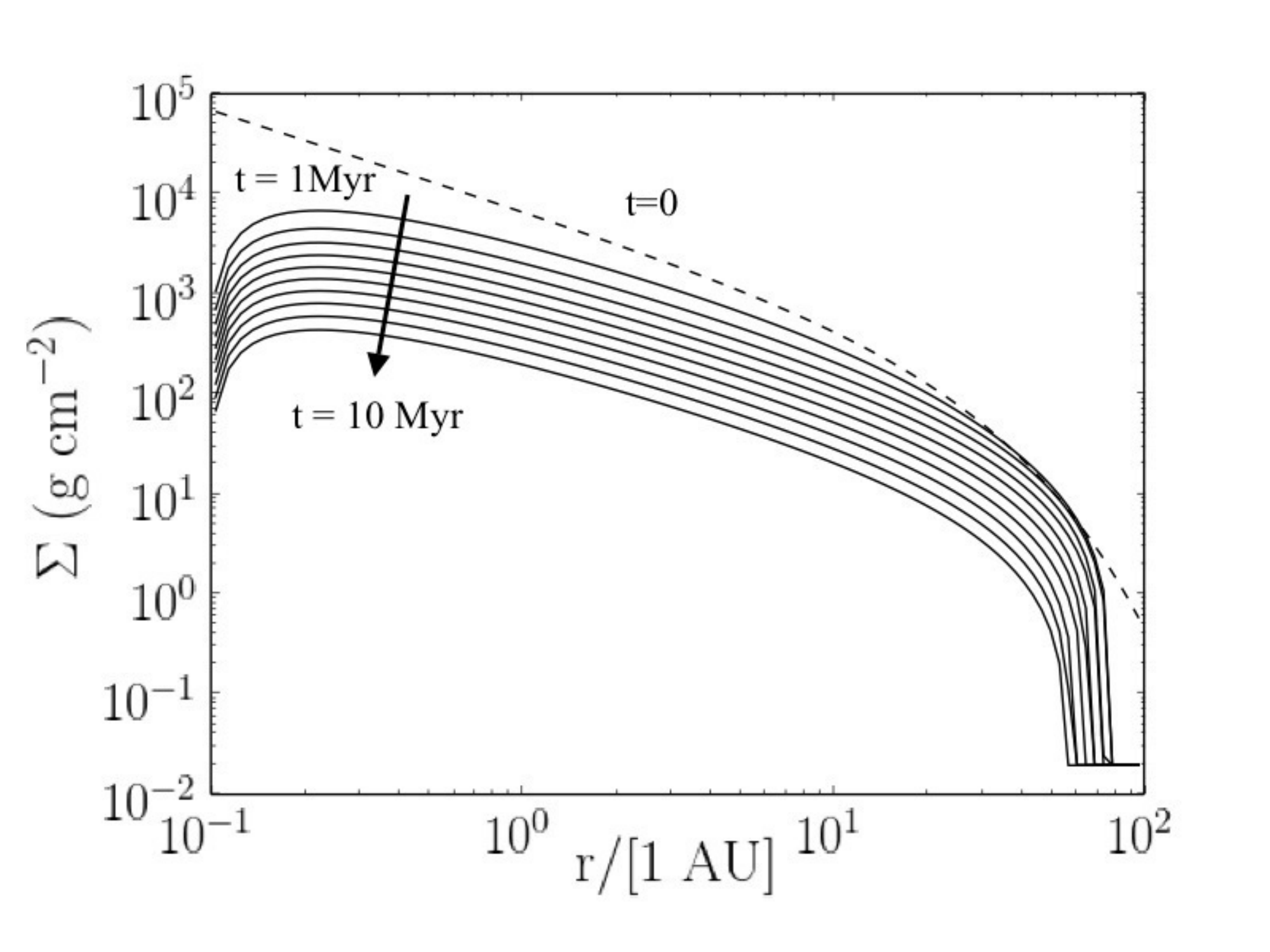}
\epsscale{0.85}
\caption{Surface density profiles $\Sigma (r,t)$ of our canonical uniform $\alpha$ case with photoevaporation. $\alpha$ is assumed to be 0.001, and $G_0 = 1000$. Each curve shows the $\Sigma$ profile at times: 0 Myr (dashed), 1 Myr, 2 Myr, .. , 10 Myr. Note that the disk is truncated to 55 AU after 10 Myr and the shape of the $\Sigma$ profile remains preserved.}
\end{figure}
\clearpage
\begin{figure}
\plotone{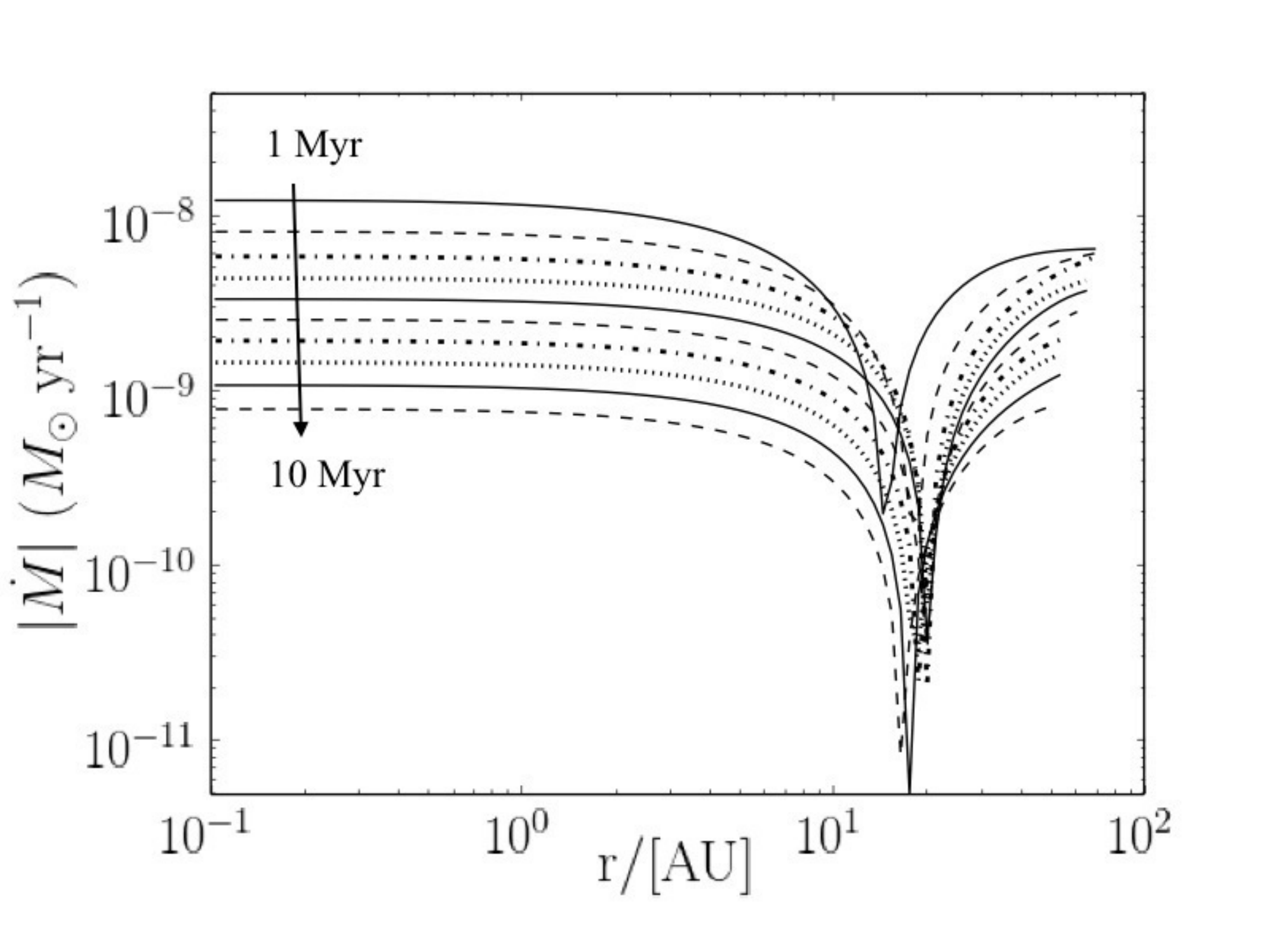}
\epsscale{0.55}
\caption{$ \rm{\dot{M}(r)}$ profiles of the disk for each successive Myr: 1 Myr, 2 Myr .. 10 Myr (solid, dashed, dot-dashed, dotted...) for the canonical uniform $\alpha$ case with photoevaporation ($G_0=1000$). Mass moves radially inward till radius = $r_T$ (transition radius) where the mass flow changes direction. Beyond $r_T$, mass flows radially outward due to photoevaporation. The dip in each curve denotes $r_T$, which is independently plotted in Fig 4.}
\end{figure}
\clearpage
\begin{figure} 
\plotone{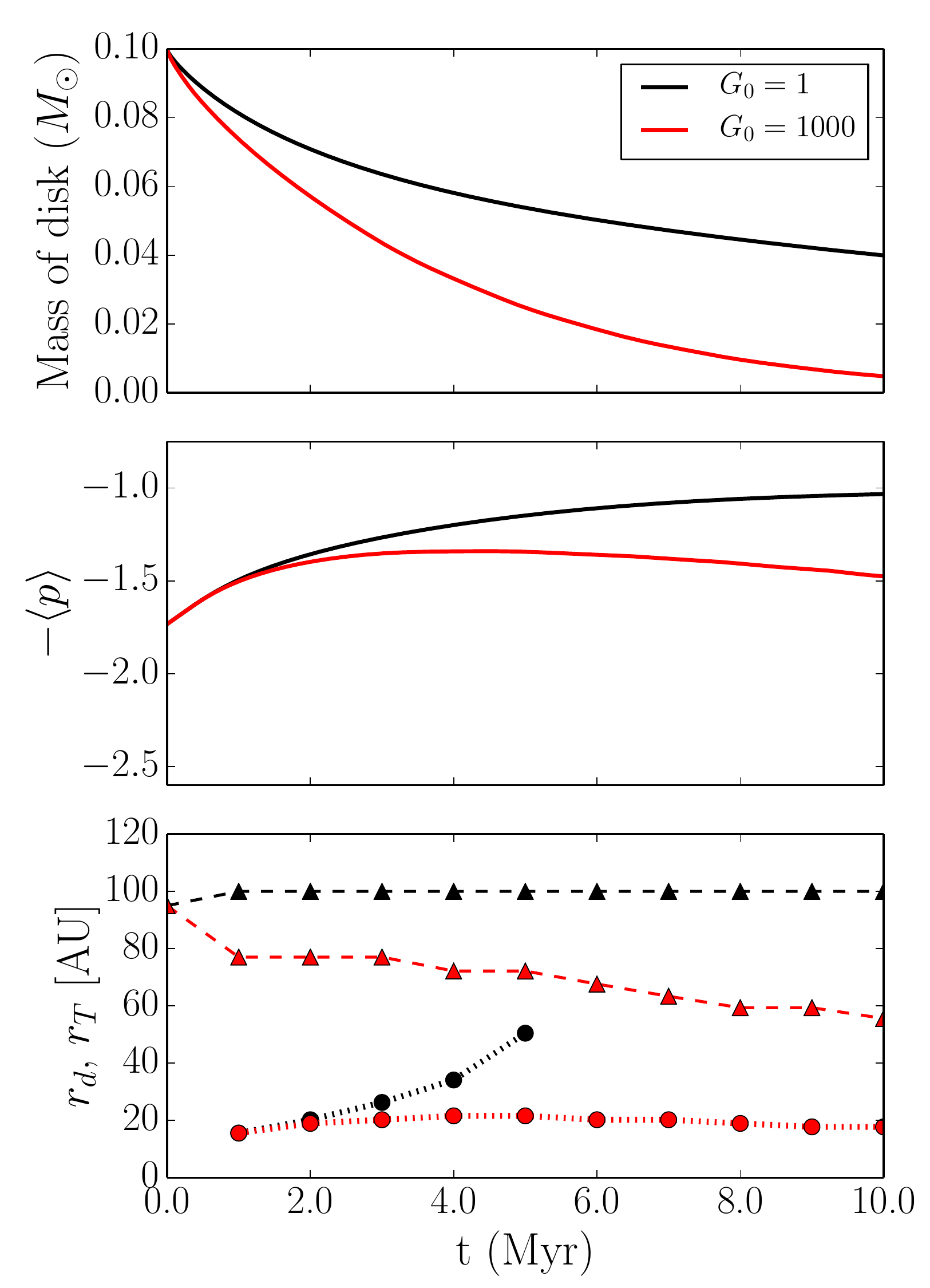}
\epsscale{0.55}
\caption{ Change in the disk properties (disk mass $M_d$, slope $\langle p \rangle$ of $\Sigma(r)$, disk outer edge $r_d$ and transition radius $r_T$) with time for the canonical uniform $\alpha$ case with photoevaporation ($G_0=1000$; red). Black curves show the non-photoevaporated case ($G_0=1$ with other parameters unchanged) for comparison. Here, $\langle p \rangle$ denotes spatial average of slope $p$ across 5-30 AU, and $r_d$ and $r_T$ are shown at each successive Myr of evolution. Non-photoevaporated case ($G_0=1$) is denoted by black triangles (for $r_d$) and black circles (for $r_T$), and photoevaporated case ($G_0=1000$) is denoted by red triangles ($r_d$) and circles ($r_T$). Note that $r_T$ moves outward with time in a non-photoevaporated disk, but moves inward with time in a photoevaporated disk after first few Myr. (For the non-photoevaporated case, $r_T$ moves beyond 100 AU after 5 Myr).}
\end{figure}
\begin{figure}
\plotone{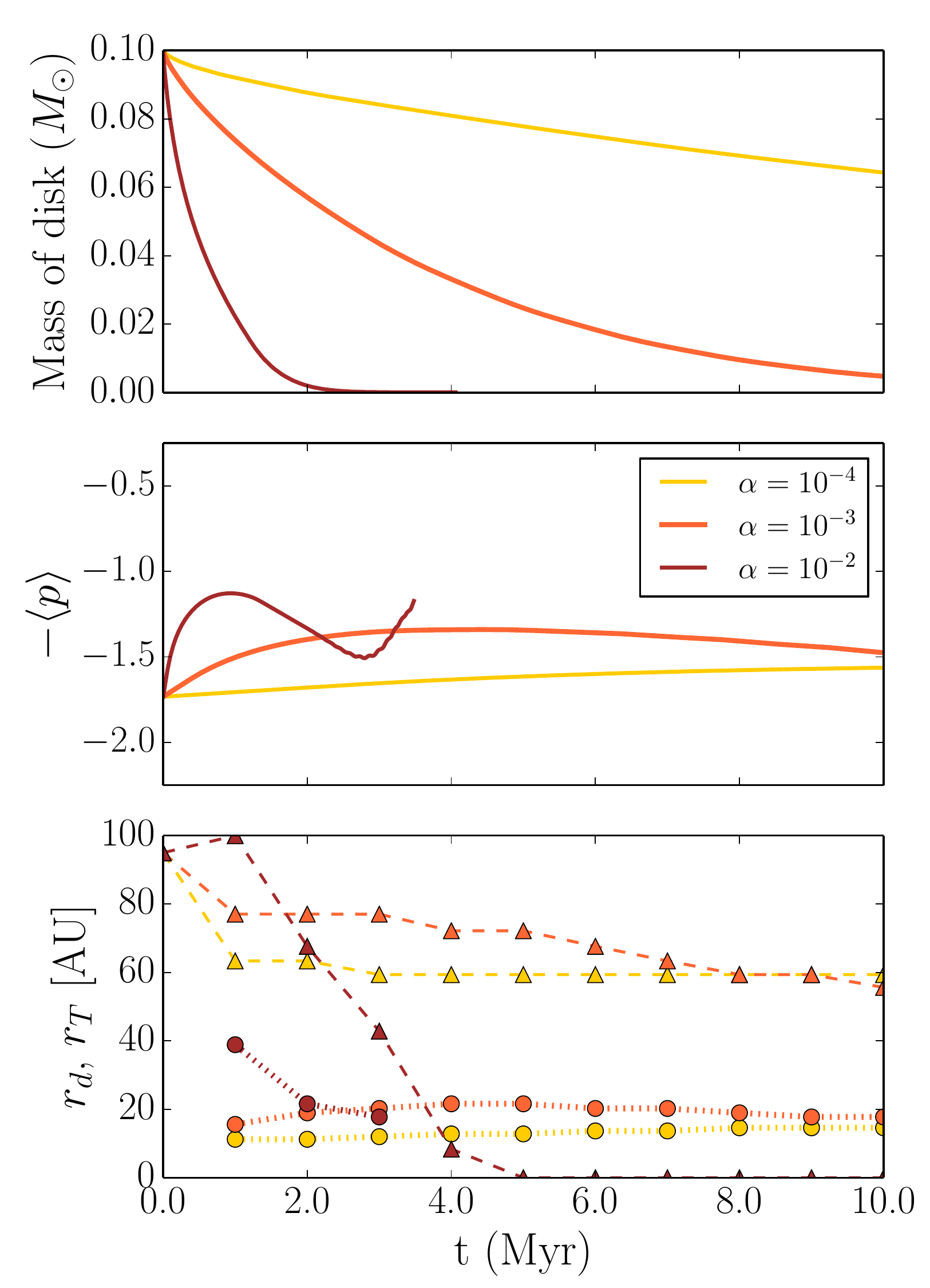}
\epsscale{0.85}
\caption{ Effect of variation of $\alpha$ on disk properties (disk mass $M_d$, slope $\langle p \rangle$ of $\Sigma(r)$, disk outer edge $r_d$ and transition radius $r_T$) with time for the canonical uniform $\alpha$ case with photoevaporation for a range of $\alpha$ values [0.01, 0.001, 0.0001]. Disk with high viscosity ($\alpha$=0.01) rapidly evolves and shrinks to $\sim$ 10 AU in 4 Myr. $\langle p \rangle$ denotes average of $p$ over 5-30 AU. Triangles denote $r_d$ points and circles denotes $r_T$ points at each Myr. (For $\alpha = 0.01$ case, the disk is too small to retain $r_T$ after 3 Myr).}
\end{figure}
\begin{figure}
\plotone{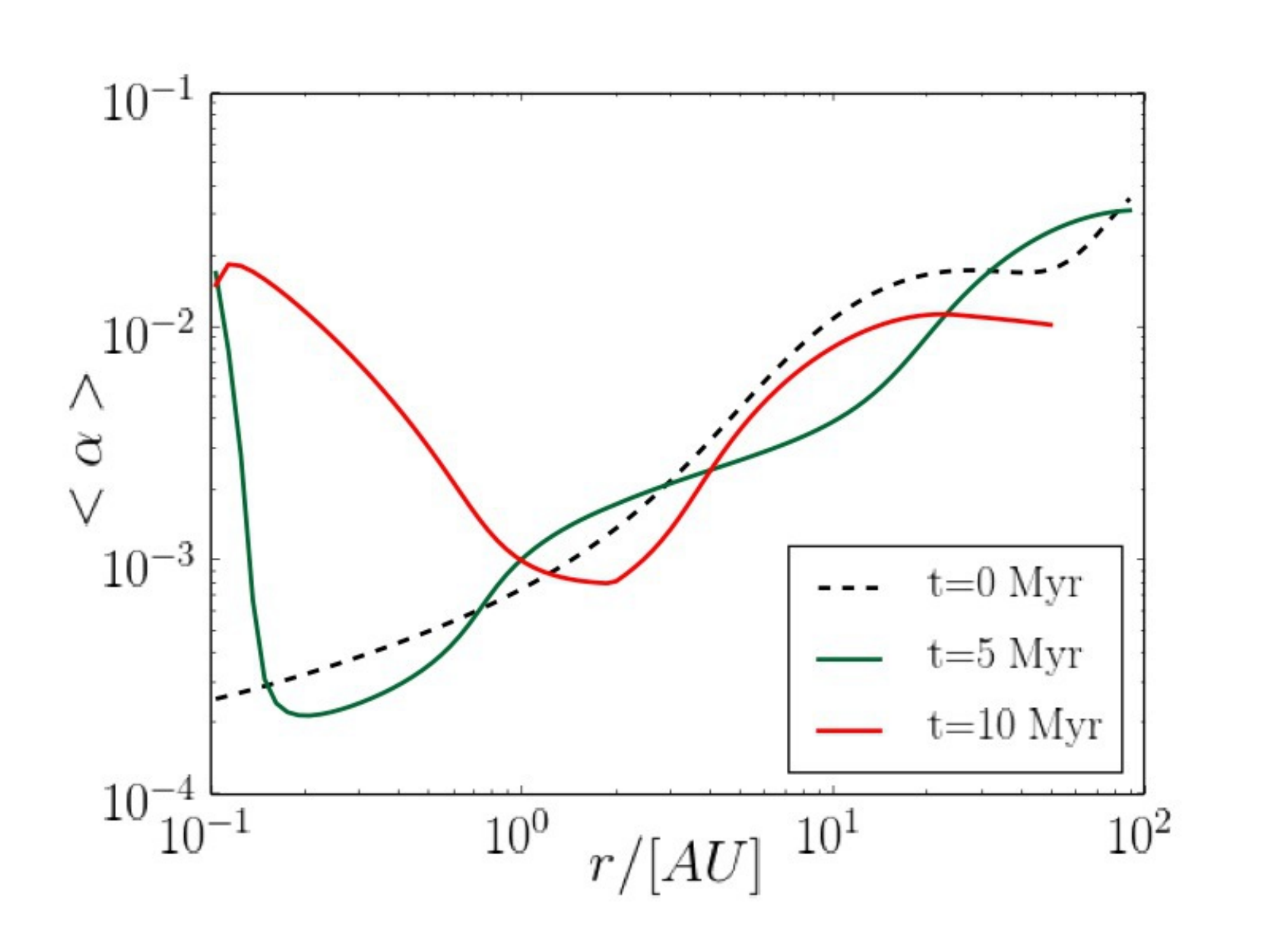}
\epsscale{0.85}
\caption{Vertically integrated and mass weighted $\langle\alpha\rangle$ as a function of $r$, at various times for the canonical computed $\alpha$ dust-free case, without photoevaporation ($G_0 = 1$). The curves are truncated at the disk radius $r_d$ at each plotted time.}
\end{figure}
\begin{figure}
\plotone{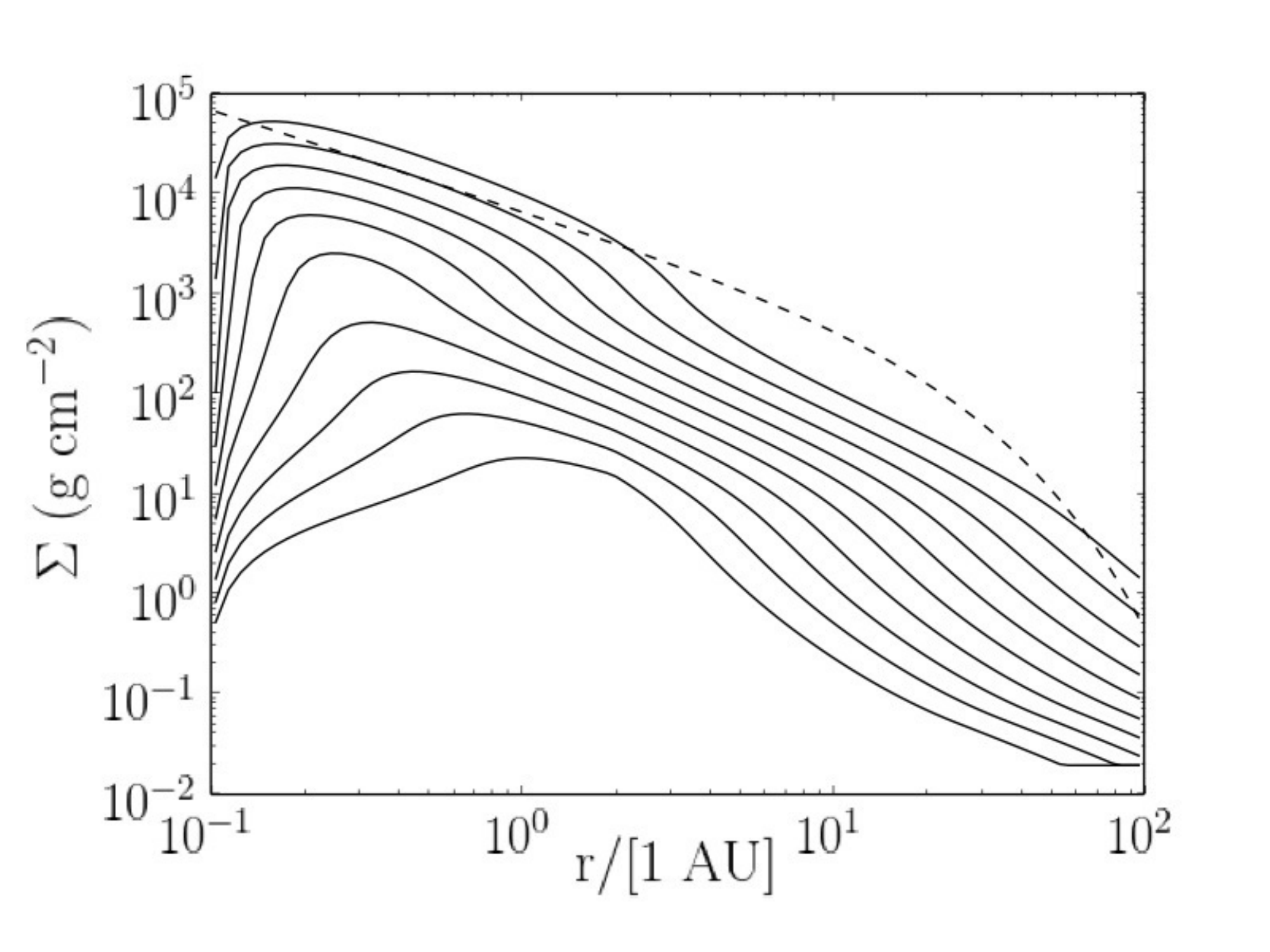}
\epsscale{0.85}
\caption{$\Sigma (r,t)$ for the canonical computed $\alpha$ case for the dust-free case without photoevaporation ($G_0 = 1$). Each curve shows the surface density profile at times 0 Myr (dashed), 1 Myr, 2 Myr, 3 Myr, .. , 10 Myr. Note the overall disk profile is maintained for several Myr.}
\end{figure}
\begin{figure}
\plotone{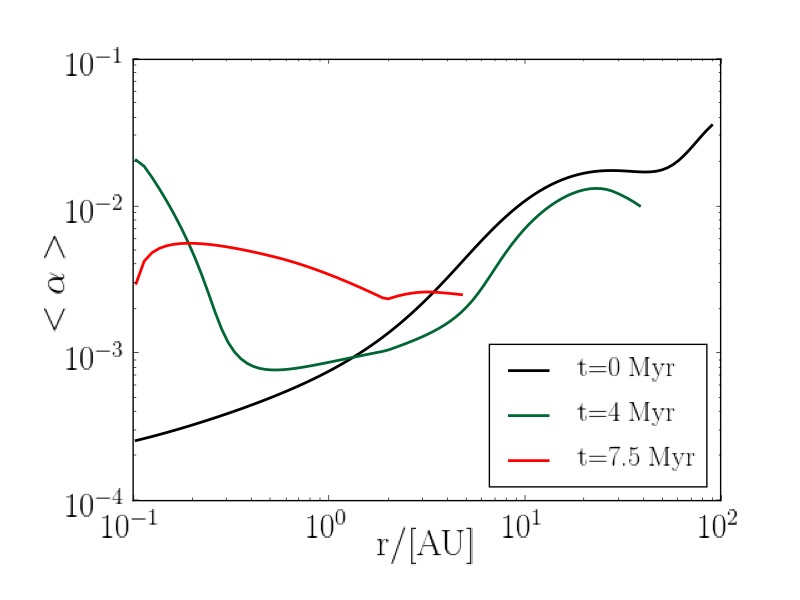}
\epsscale{0.85}
\caption{Vertically integrated and mass weighted $\langle\alpha\rangle$ as a function of $r$, at various times for the canonical computed $\alpha$ dust-free case with photoevaporation ($G_0=1000$). The curves are truncated at the disk radius $r_d$ at each plotted time.}
\end{figure}
\clearpage
\begin{figure}
\plotone{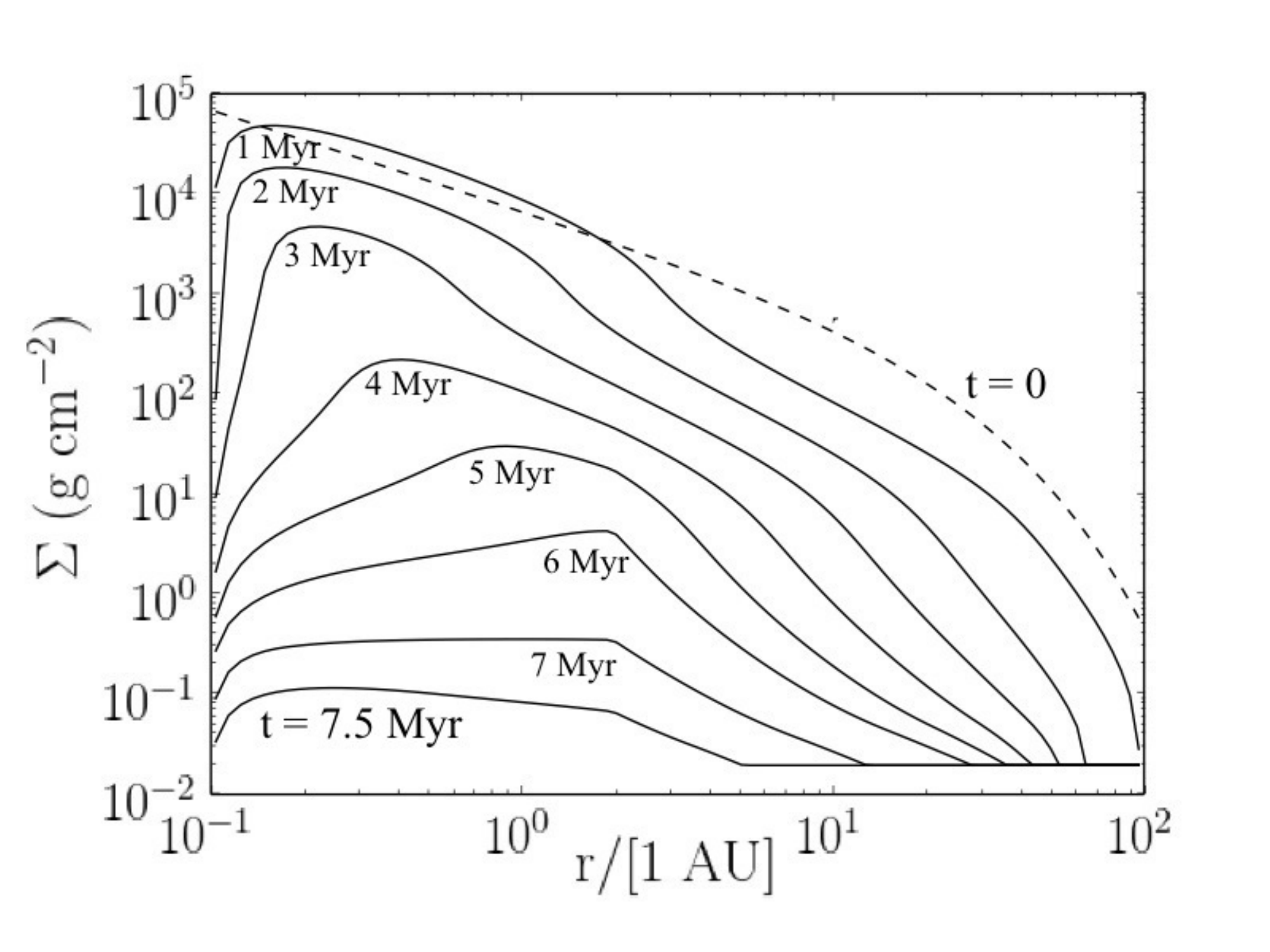}
\epsscale{0.85}
\caption{$\Sigma (r,t)$ for the canonical computed $\alpha$ case for the dust-free case with photoevaporation ($G_0 = 1000$). Each curve shows the surface density profile at times 0 Myr (dashed), 1 Myr, 2 Myr, .. , $t_{\rm{final}}$. $t_{\rm{final}} \sim 7.5$ Myr. Note that the disk rapidly shrinks to 5 AU within 7.5 Myr.}
\end{figure}
\clearpage
\begin{figure}
\plotone{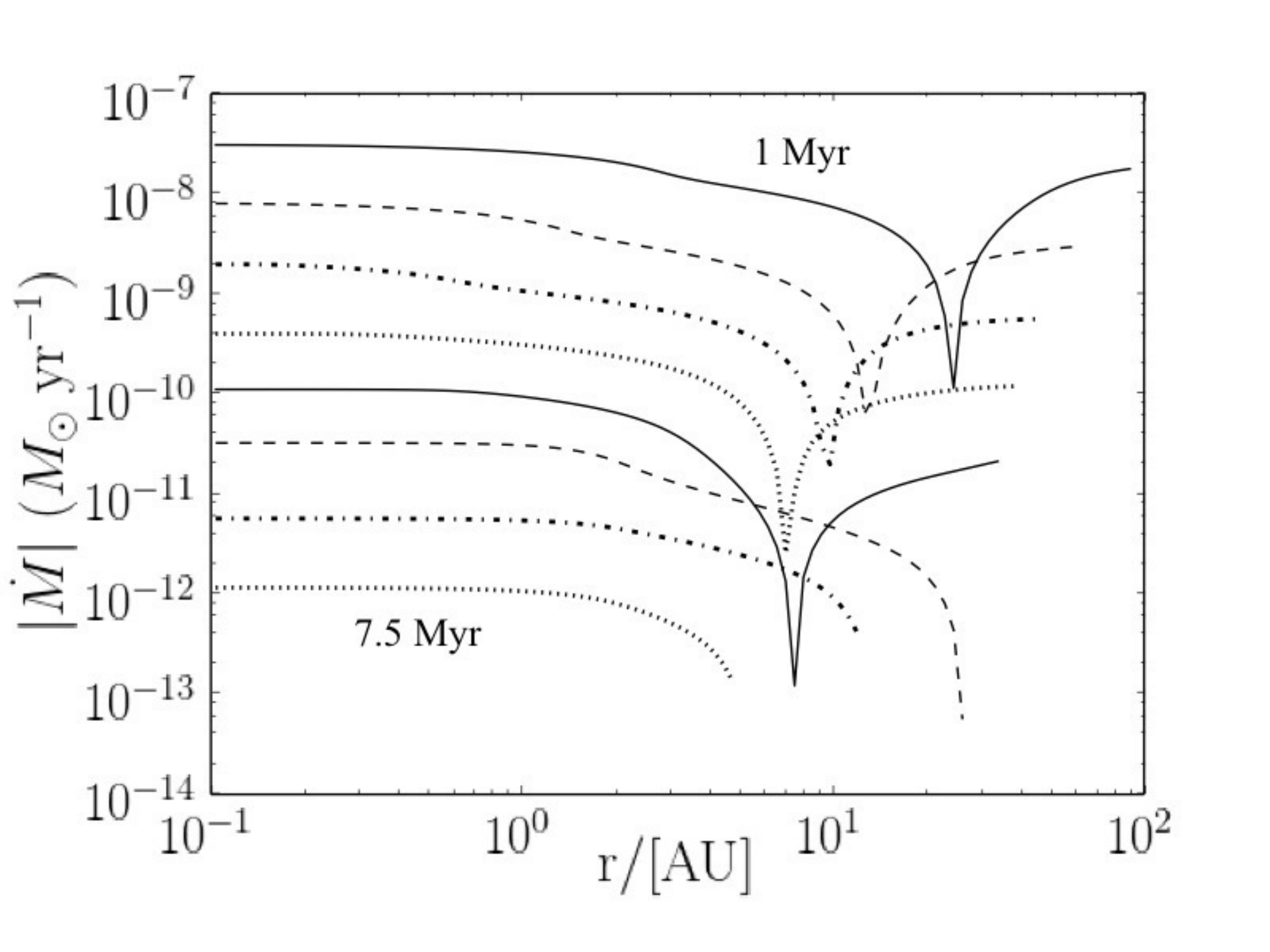}
\epsscale{0.65}
\caption{$\dot{M}$ profiles of the disk for each successive Myr: 1Myr, 2 Myr, .. , 10 Myr (solid, dashed, dot-dashed, dotted..) for the canonical computed dust-free $\alpha$ case with photoevaporation ($G_0=1000$). The dip in each curve denotes $r_T$, which is independently plotted in Fig 11.}
\end{figure}
\clearpage
\begin{figure}
\plotone{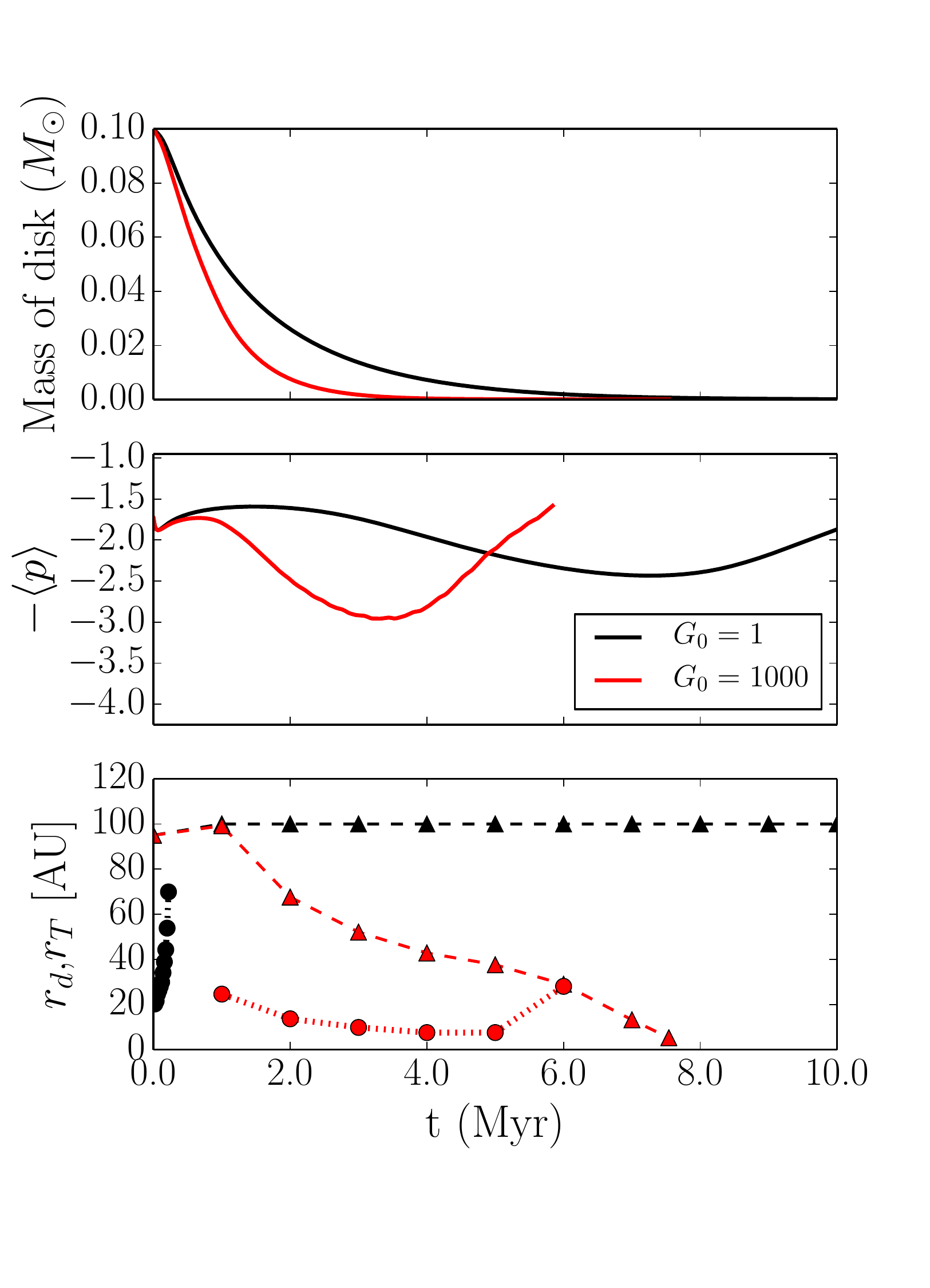}
\epsscale{0.85}
\caption{ Change in the disk properties (disk mass $M_d$, slope $\langle p \rangle$ of $\Sigma(r)$, disk outer edge $r_d$ and transition radius $r_T$) with time for the canonical varying $\alpha$ dust-free case with photoevaporation ($G_0=1000$; red). Black curves show the non-photoevaporated case ($G_0=1$ with other parameters unchanged) for comparison. Same as in Fig.\ 4. $r_T$ for the non-photoevaporated case moves beyond 100 AU within 1 Myr.}
\end{figure}
\begin{figure}
\plotone{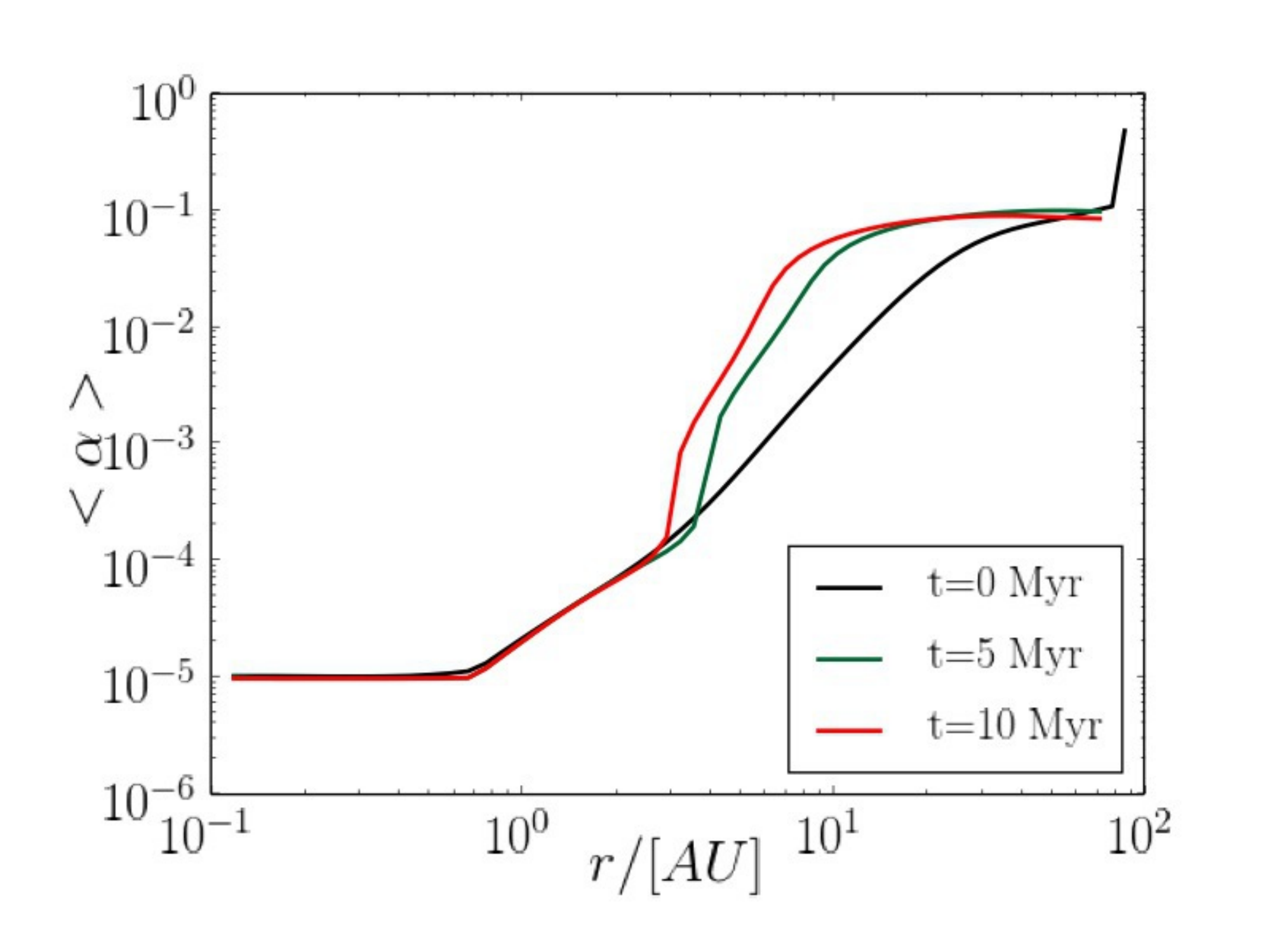}
\epsscale{0.85}
\caption{Vertically integrated and mass weighted $\langle\alpha\rangle$ as a function of $r$, at various times for the canonical computed $\alpha$ case with dust (without photoevaporation; $G_0=1$)}
\end{figure}
\clearpage
\begin{figure}
\plotone{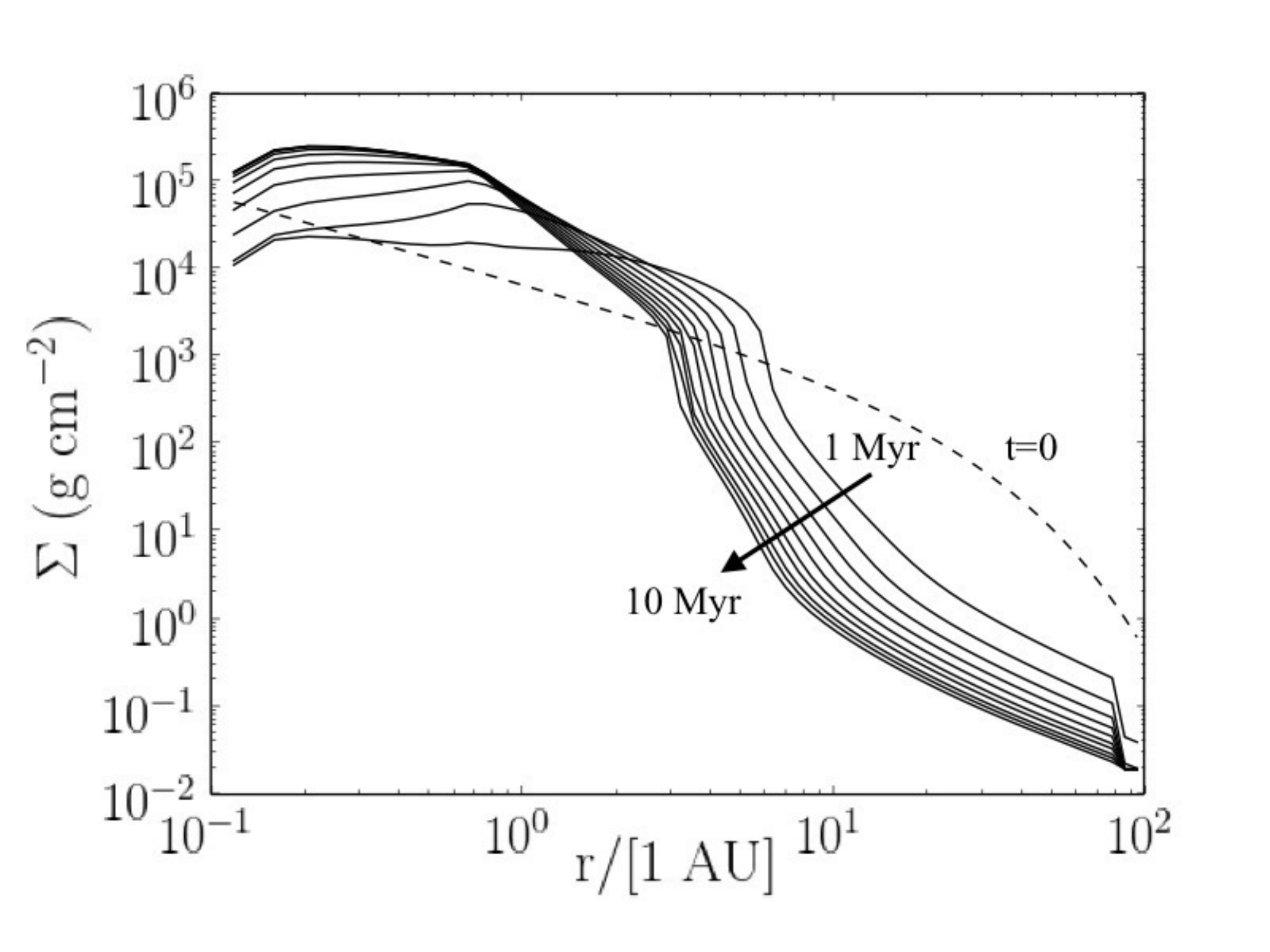}
\epsscale{0.85}
\caption{$\Sigma (r,t)$ for the canonical computed $\alpha$ case with dust for a non-photoevaporated disk ($G_0 = 1$). Each curve shows the surface density profile at times 0 Myr (dashed), 1 Myr, 2 Myr .., 10 Myr. Note that the dust stalls the inner disk evolution and there is a large-scale distribution of mass towards the inner and mid-disk.}
\end{figure}
\begin{figure}
\plotone{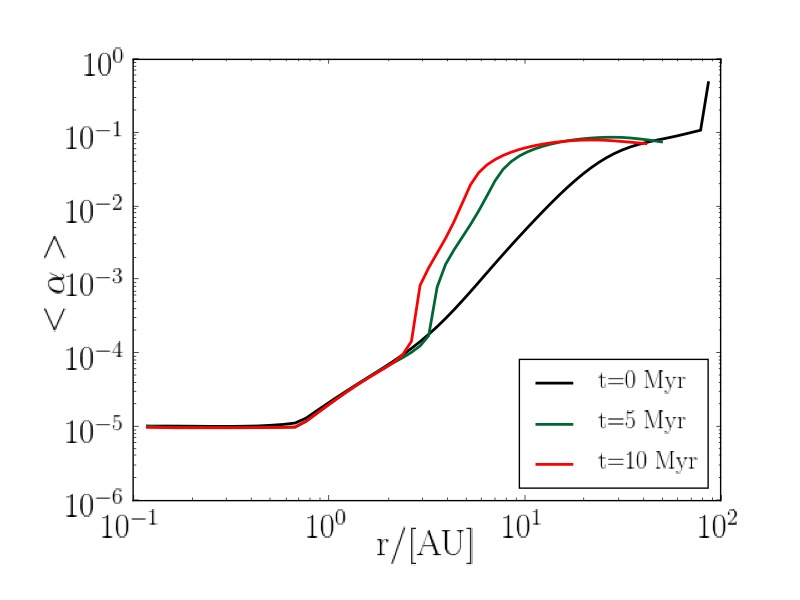}
\epsscale{0.85}
\caption{Vertically integrated and mass weighted $\langle\alpha\rangle$ as a function of $r$, at various times for the canonical computed $\alpha$ case with dust (with photoevaporation; $G_0=1000$).}
\end{figure}
\clearpage
\begin{figure}
\plotone{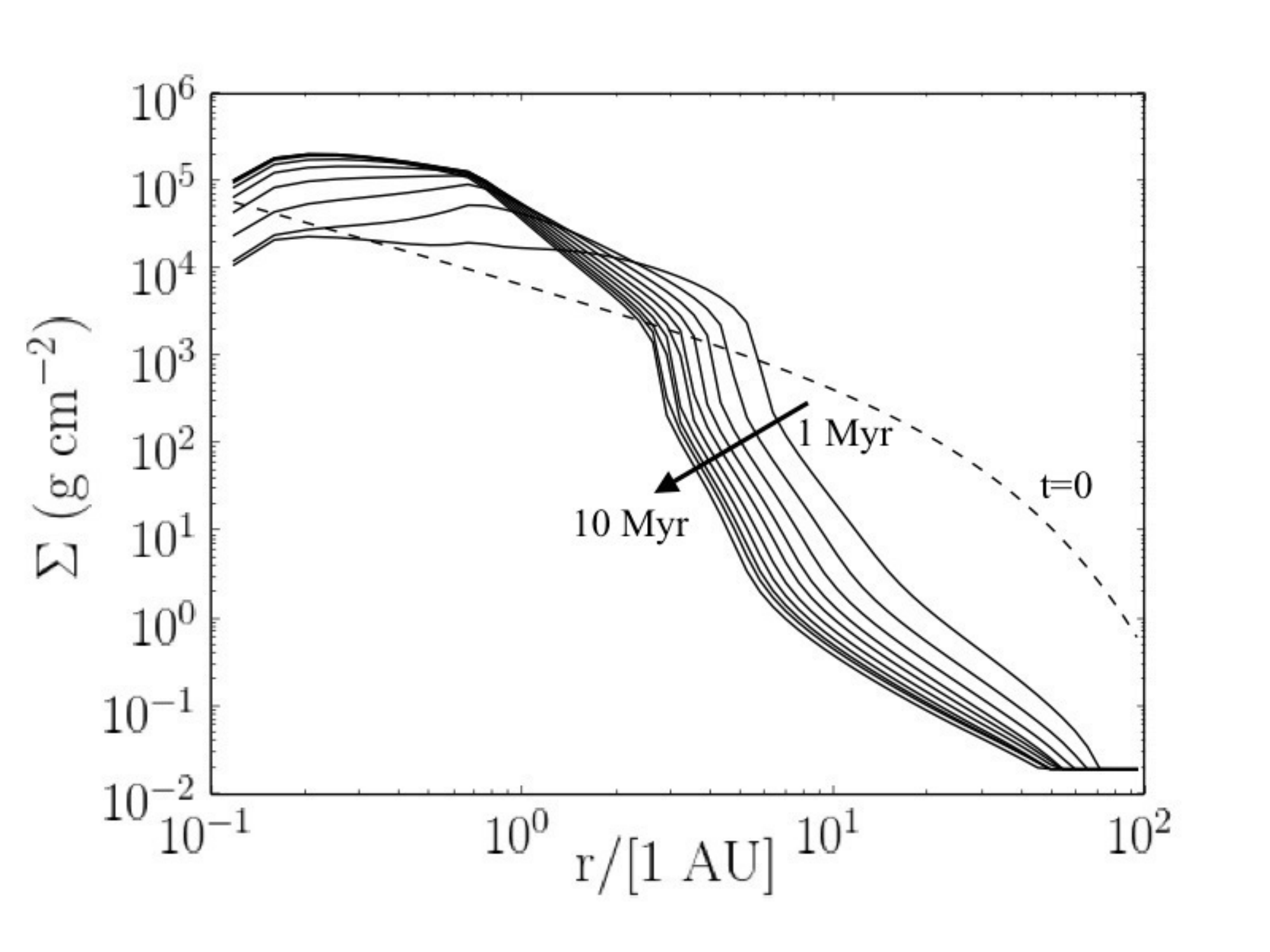}
\epsscale{0.85}
\caption{$\Sigma (r,t)$ for the canonical computed $\alpha$ case with dust for a photoevaporated disk ($G_0 = 1000$). Each curve shows the surface density profile at times 0 Myr (dashed), 1 Myr, 2 Myr, .. , 10 Myr. Note that the dust stalls the inner disk evolution and there is a large-scale distribution of mass towards the inner and mid-disk, and the disk is truncated to $\sim$ 50 AU within 10 Myr.}
\end{figure}
\clearpage
\begin{figure}
\plotone{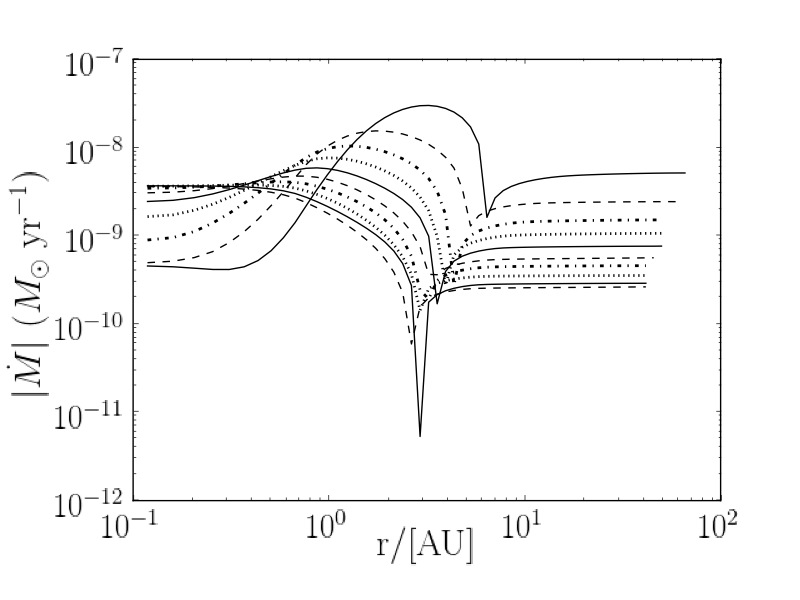}
\epsscale{0.65}
\caption{This plot shows the $\dot{M}$ profiles of the disk for each successive Myr: 1Myr, 2 Myr, .. , 10 Myr (solid, dashed, dot-dashed, dotted...) for the canonical computed $\alpha$ case with dust and photoevaporation ($G_0=1000$). Dips in each curve represent $r_T$, where mass flow in the disk changes direction from inward to outward. Note how $r_T$ moves inward with time (independently plotted in Fig.\ 17).}
\end{figure}
\clearpage
\begin{figure}
\plotone{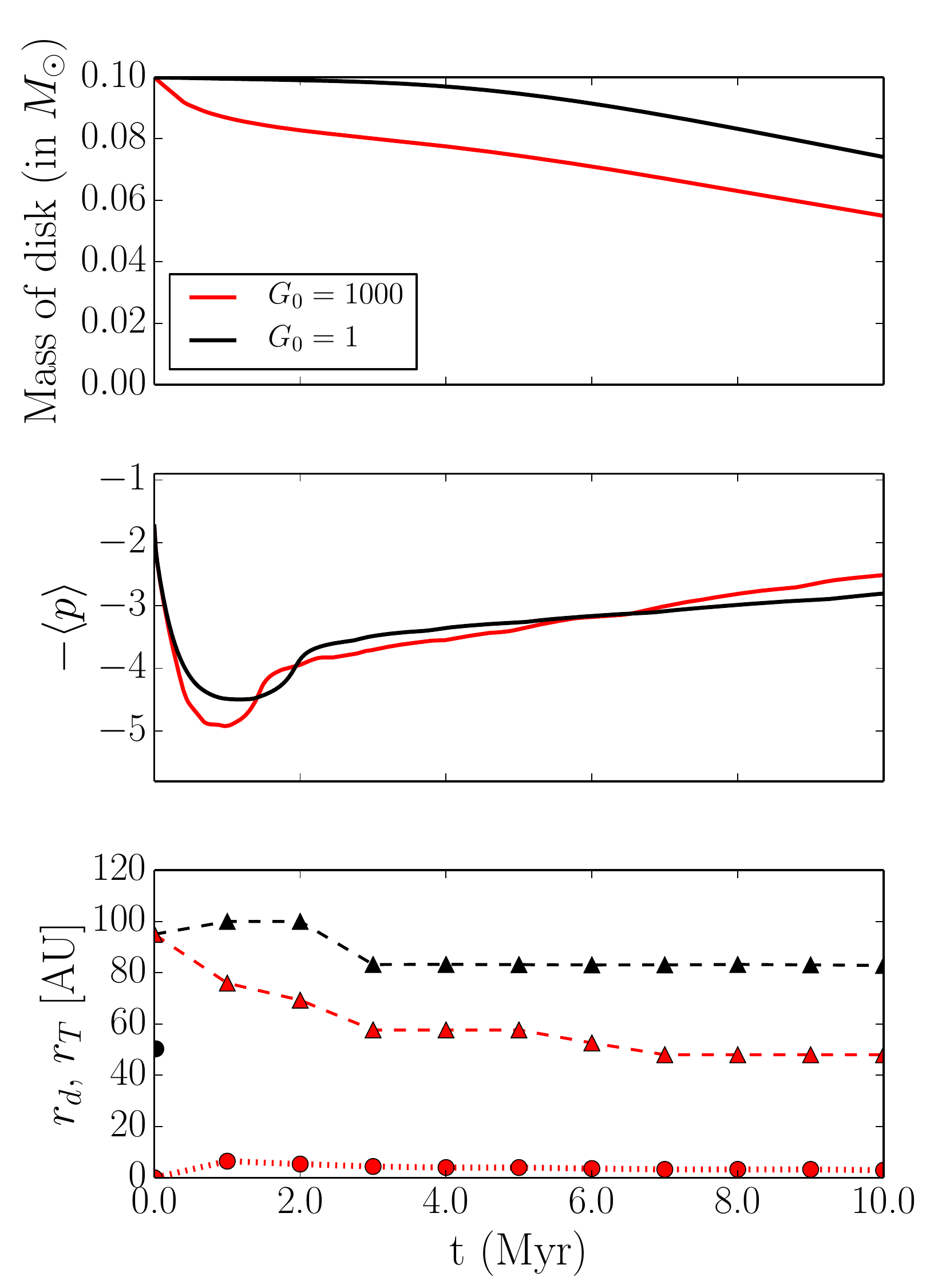}
\epsscale{0.65}
\caption{Change in the disk properties (disk mass $M_d$, slope $\langle p \rangle$ of $\Sigma(r)$, disk outer edge $r_d$ and transition radius $r_T$) with time for the canonical varying $\alpha$ case with dust, and subjected to photoevaporation ($G_0=1000$; red). Black curves show the non-photoevaporated case ($G_0=1$ with other parameters unchanged) for comparison. Same as Figs.\ 4 and 11. $r_T$ moves beyond 100 AU in $\ll$ 1 Myr (i.e., in 20000 yr) for the non-photoevaporated case.}
\end{figure}
\begin{figure}
\plotone{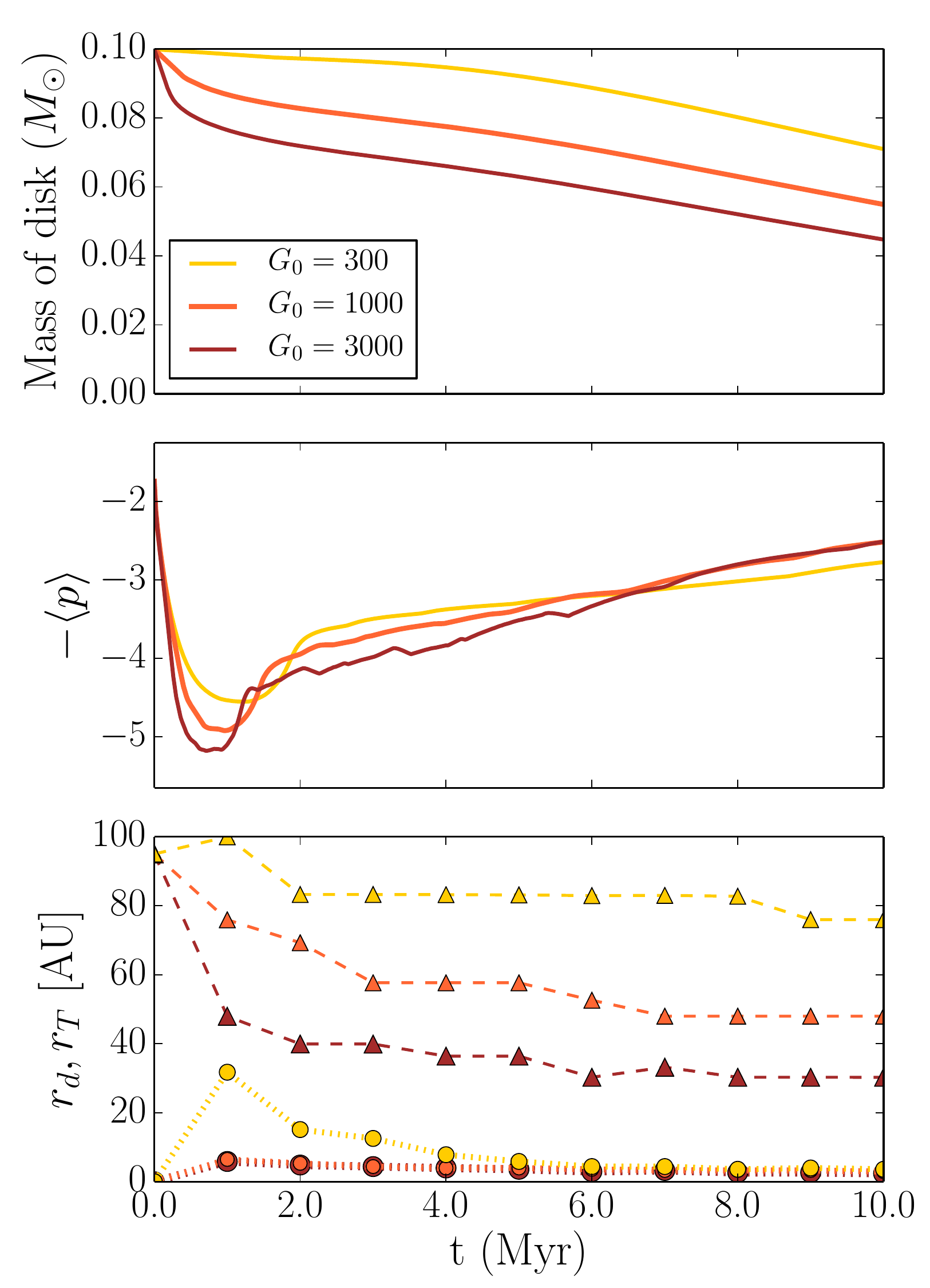}
\epsscale{0.65}
\caption{Effect of variation of $G_0$ on disk properties (disk mass $M_d$, slope $\langle p \rangle$ of $\Sigma(r)$, disk outer edge $r_d$ and transition radius $r_T$) with time for the canonical computed $\alpha$ case with dust and subjected to different FUV fluxes ($G_0$ = [300, 1000, 3000]). Same as Fig.\ 5.}
\end{figure}
\clearpage
\begin{figure}
\plotone{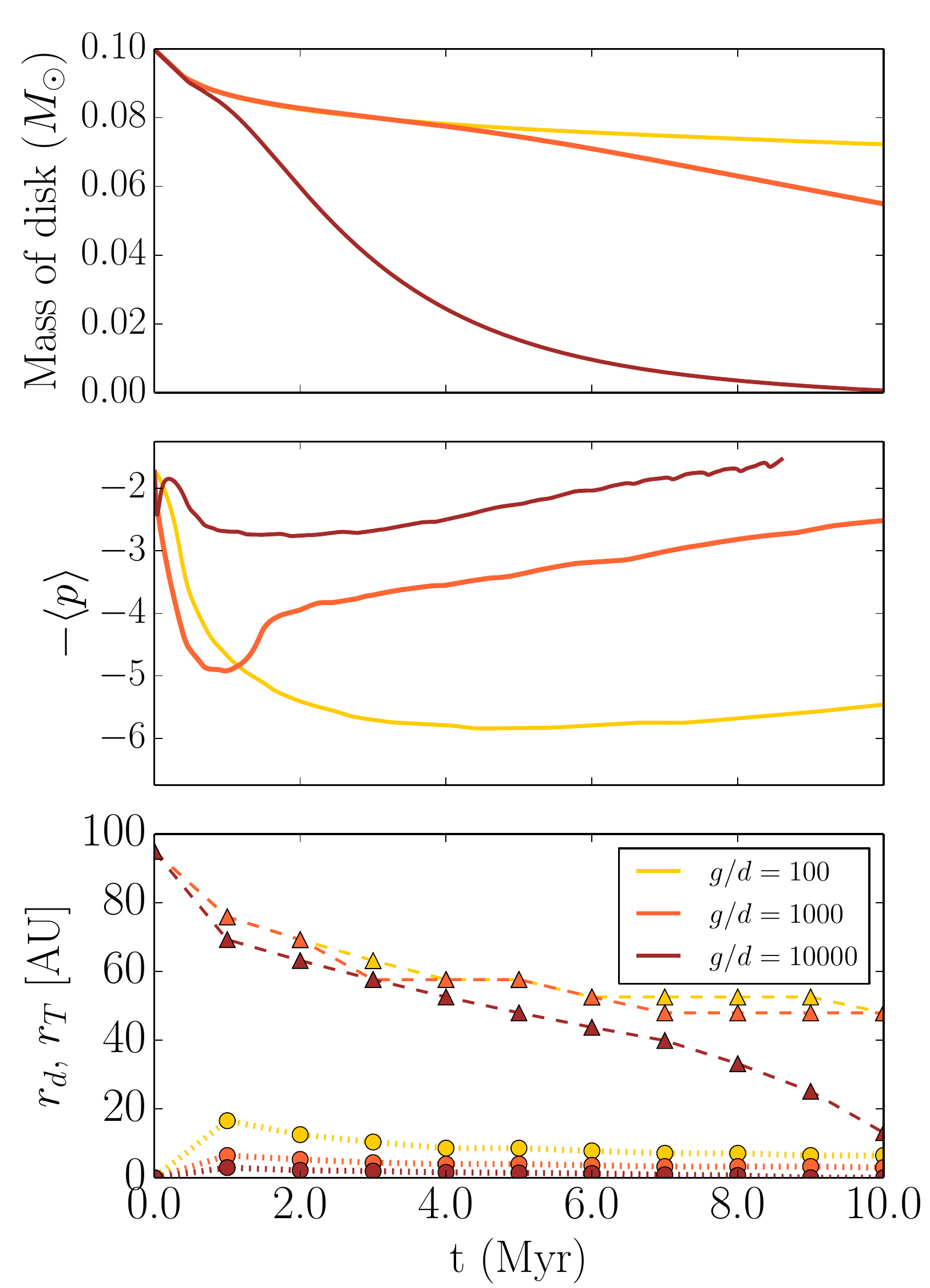}
\epsscale{0.65}
\caption{Effect of variation of gas-to-dust (g/d) ratio on disk properties (disk mass $M_d$, slope $\langle p \rangle$ of $\Sigma(r)$, disk outer edge $r_d$ and transition radius $r_T$) with time for the canonical computed $\alpha$ case with dust and subjected to photoevaporation ($G_0=1000$). A range of (g/d) [100, 1000, 10000] was explored. Same as Fig.\ 5.}
\end{figure}
\clearpage
\begin{figure}
\plotone{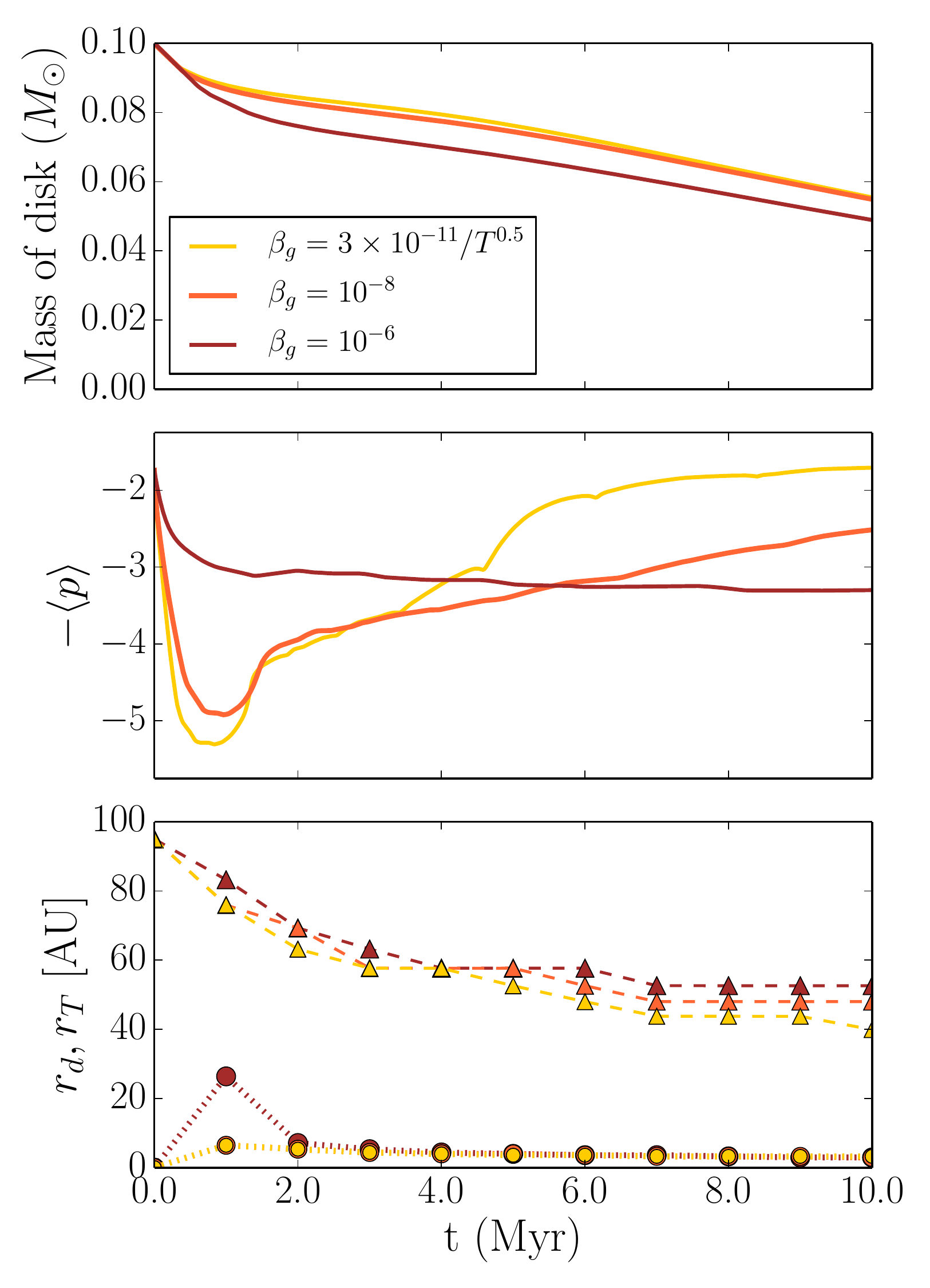}
\epsscale{0.65}
\caption{Effect of variation of the gas phase recombination coefficient $\beta_g$ on disk properties (disk mass $M_d$, slope $\langle p \rangle$ of $\Sigma(r)$, disk outer edge $r_d$ and transition radius $r_T$) with time for the canonical computed $\alpha$ case with dust and subjected to photoevaporation ($G_0=1000$). A range of $\beta_g$ ($10^{-6} \,{\rm cm}^{3}\, {\rm s}^{-1}$ for molecular ion dominated chemistry, $10^{-11} \,{\rm cm}^{3}\, {\rm s}^{-1}$ for metal ion dominated chemistry as well as an intermediate value $10^{-8} \,{\rm cm}^{3}\, {\rm s}^{-1}$ accounting for chemistry that is driven by both species) was explored. Same as Fig.\ 5}
\end{figure}
\clearpage
\begin{figure}
\plotone{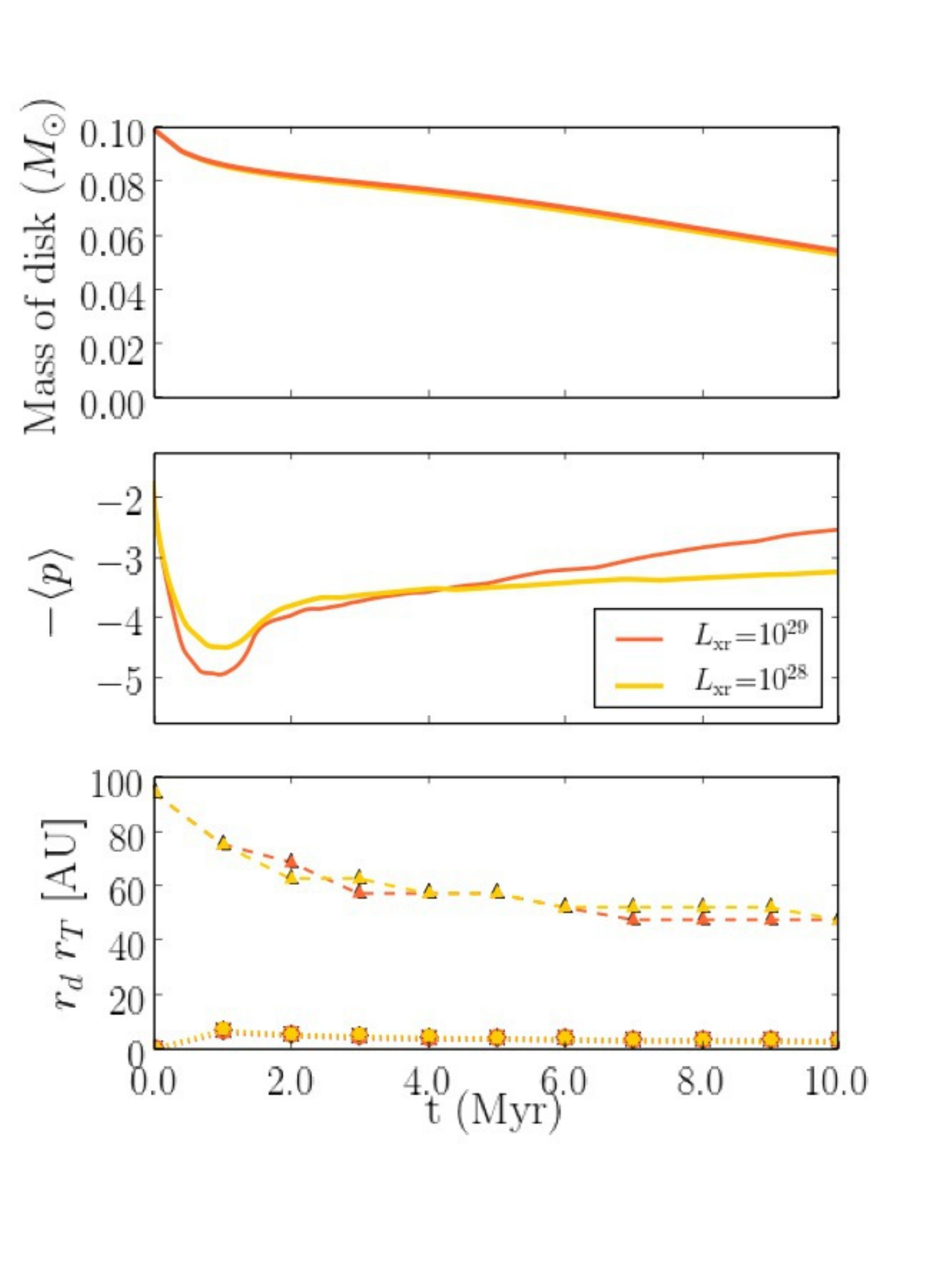}
\epsscale{0.65}
\caption{Effect of variation of $\rm{L_{xr}}$ on disk properties (disk mass $M_d$, slope $\langle p \rangle$ of $\Sigma(r)$, disk outer edge $r_d$ and transition radius $r_T$) with time for the canonical computed $\alpha$ case with dust and subjected to photoevaporation ($G_0=1000$). $\rm{L_{xr}}$ of $10^{28}$ and $10^{29} \,{\rm ergs\,s}^{-1}$ were explored. Same as Fig.\ 5.}
\end{figure}
\clearpage
\begin{figure}
\plotone{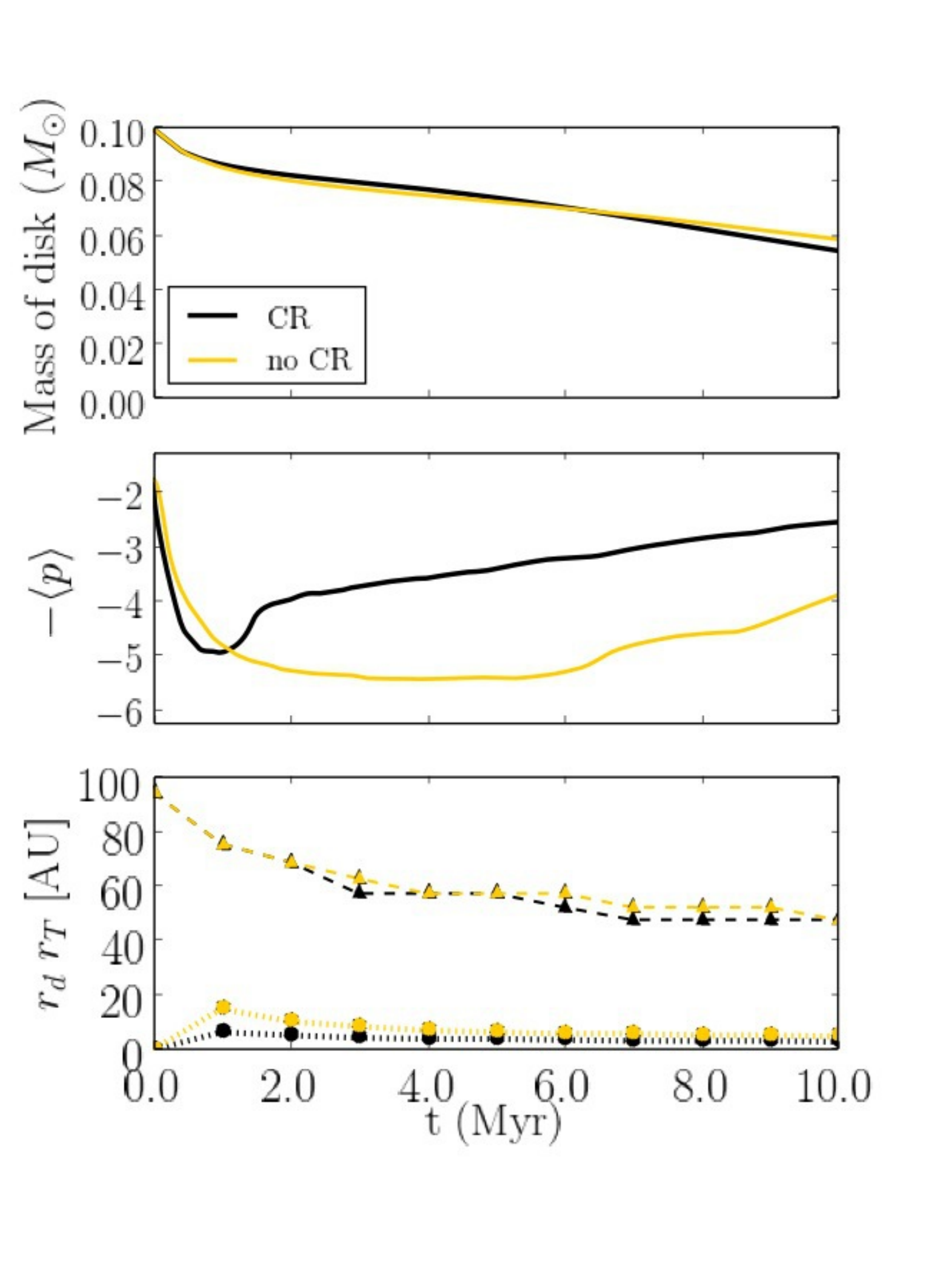}
\epsscale{0.65}
\caption{Effect of exclusion of cosmic rays (CR) on disk properties (disk mass $M_d$, slope $\langle p \rangle$ of $\Sigma(r)$, disk outer edge $r_d$ and transition radius $r_T$) with time for the canonical computed $\alpha$ case with dust and subjected to photoevaporation ($G_0=1000$). Same as Fig.\ 5.}
\end{figure}

\clearpage
\begin{figure}
\plotone{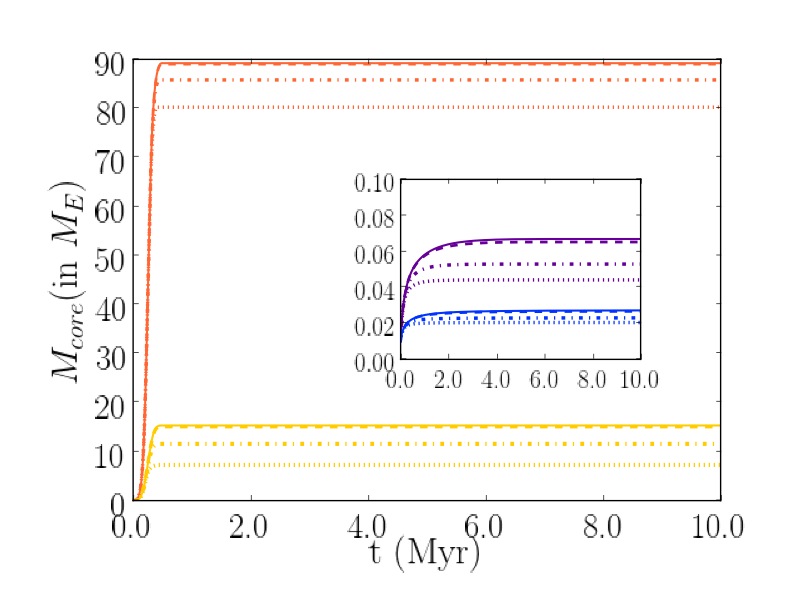}
\epsscale{0.65}
\caption{Effect of photoevaporation on the growth of core masses with time. The four planet cores are represented by color (orange: Jupiter, yellow: Saturn, violet: Neptune (inset), blue: Uranus (inset). Different photo evaporative FUV fluxes are represented by solid, dashed, dashed-dotted and dotted lines for G0 = 1, 300, 1000, 3000 respectively. The inset axes labels are the same as that of the plot axes.}
\end{figure}

\end{document}